\documentclass[twocolumn,showpacs,prb,a4paper,floatfix]{revtex4-1}
\usepackage{amssymb}
\usepackage{graphicx}
\usepackage{amsmath}
\usepackage{graphicx}
\usepackage{dcolumn}
\usepackage{bm}
\usepackage{color}

\newcommand{\bi}{\begin{itemize}}
\newcommand{\ei}{\end{itemize}}
\newcommand{\be}{\begin{equation}}
\newcommand{\ee}{\end{equation}}
\newcommand{\bea}{\begin{eqnarray}}
\newcommand{\eea}{\end{eqnarray}}

\begin{document}

\title{ Optical conductivity of disordered graphene beyond the
Dirac cone approximation}
\author{Shengjun Yuan$^{1}$, Rafael Rold\'{a}n$^{1,2}$, Hans De Raedt$^{3}$,
and Mikhail I. Katsnelson$^1$}
\date{\today }

\begin{abstract}
In this paper we systemically study the optical conductivity and
density of states of disorded graphene beyond the Dirac cone
approximation. The optical conductivity of graphene is computed by
using the Kubo formula, within the framework of a full $\pi $-band
tight-binding model. Different types of non-correlated and
correlated disorders are considered, such as random or Gaussian
potentials, random or Gaussian nearest-neighbor hopping
parameters, randomly distributed vacancies or their clusters, and
random adsorbed hydrogen atoms or their clusters. For a large
enough concentration of resonant impurities, a new peak in the
optical conductivity is found, associated to transitions between
the midgap states and the Van Hove singularities of the main
$\pi$-band. We further discuss the effect of doping on the
spectrum, and find that small amounts of resonant impurities are
enough to obtain a background contribution to the conductivity in
the infra-red part of the spectrum, in agreement with recent
experiments.
\end{abstract}


\affiliation{\centerline{$^1$Institute for Molecules and Materials, Radboud University of Nijmegen, NL-6525AJ Nijmegen, The Netherlands}\\
\centerline{$^2$Instituto de Ciencia de Materiales de Madrid, CSIC, Cantoblanco E28049 Madrid, Spain}
\centerline{$^3$Department of Applied Physics, Zernike Institute for Advanced Materials,}
\centerline{University of Groningen, Nijenborgh 4, NL-9747AG Groningen, The Netherlands}}

\maketitle

\section{Introduction}

An important part of our knowledge on the electronic properties
of graphene, which consist of a two-dimensional (2D) lattice of carbon atoms,%
\cite{CG09} can be deduced from optical spectroscopy measurements
(for recent reviews see Refs. \onlinecite{P10,OP10}). Infrared
spectroscopy experiments allows for the control of interband excitations by
means of electrical gating,\cite{WangF2008,LB08} similarly as
electrical transport in field effect transistors. Within the
simplest Dirac cone approximation, only vertical in wave-vector
space transitions across the Dirac point are optically active,
leading to a constant value for the optical conductivity of
undoped graphene of $\sigma _{0}=\pi e^{2}/2h$. This leads
to a frequency-independent absorption of $\pi \alpha \approx 2.3\%$, where $%
\alpha=e^2/\hbar c\approx 1/137 $ is the fine structure constant.
This fact was observed for suspended graphene in
experiments in the visible range of the spectrum\cite{Nair2008}
and it was later confirmed by further experiments in suspended
graphene\cite{Mak2008,Fei2008} and epitaxial graphene on SiC
substrate.\cite{Dawlaty2008} For doped graphene with nonzero
chemical potential $\mu $, at zero temperature, in the absence of
disorder and without considering many body effects, the allowed
excitations are only those between particle-hole pairs with an
energy difference larger than $2\mu $, due to Pauli's exclusion
principle. This would lead to a zero infrared conductivity below
the energy $\omega=2\mu $, and the optical conductivity should be
simply a step function $\sigma \left( \omega \right) =\sigma
_{0}\Theta \left( \omega -2\mu \right) $. However, a background
contribution to the optical conductivity between $0<\omega<2\mu$
was observed in Refs. \onlinecite{LB08,Mak2008}, pointing out the
relevance of disorder and many body effects. 
Another characteristic of the optical
spectrum is the Drude peak, which is built from a transfer of
spectral weight from the low-energy
interband conductance to the $\omega\rightarrow 0$ region of the spectrum,%
\cite{Kuzmenko2008} although
a strong suppression of the Drude peak at infrared energies
has recently been observed.\cite{Horng2011}
Furthermore, the flattening of the $\pi$-bands at energies away from the
Dirac point is responsible for the strong peak in the spectrum at higher energies (of
the order of 5eV) which is associated to optical transitions between states
of the Van Hove singularities.\cite{Fei2008,MSH11,SR11} Finally, a method to
control the intermediate excited states in inelastic light scattering
experiments has been also reported, revealing the important role of quantum
interference in Raman scattering.\cite{ChenCF2011}

This intense experimental work has been accompanied by a series of
theoretical studies which have treated the problem of the optical
conductivity at different levels of approximation.\cite%
{Ando2002,Gruneis2003,Peres2006,Gusynin2006,SPG07,Gusynin2007,Stauber2008,Stauber2008b,MinHK2009}
For example, it has been suggested that the presence of spectral
weight in the \textit{forbidden} region of the optical spectrum of
doped graphene (below $\omega=2\mu$) can be associated to
disorder,\cite{Ando2002,Peres2008}
electron-electron interaction\cite{Grushin2009} or excitonic effects.\cite%
{Peres2010} In particular, the effect of electron interaction in
the spectrum has been considered in Refs.
\onlinecite{Mikhailov2007,Mishchenko2007,Falkovsky2007,Falkovsky2007b,Katsnelson2008,RLG08,Sheehy2009,Juricic2010,Giuliani2011}.
Furthermore, understanding the role played by the different kinds
of disorder that can be present in this material is essential to
increase the
mobility of the samples. Besides the long-range charged impurities,\cite%
{Ando2007,Stauber2008b,Juan2010} other possible scattering sources
such as ripples,\cite{KG08} strong random on-site potentials,\cite%
{YRK10} large concentration of hydrogen adatoms,\cite{YRK10} strain \cite%
{Pereira2010,Pellegrino2010} or random deformations of the honeycomb lattice have been considered.\cite{CV09,Sinner2011}

In this paper, we perform a systemic study of the optical spectrum
of graphene with different kinds of disorder for both doped and
undoped graphene, such as the randomness of the on-site potentials
and fluctuation of the nearest-neighbor hopping. Special attention
is paid to the presence of resonant impurities, e.g., vacancies
and hydrogen adatoms, which have been proposed as the main factor
limiting the carrier mobility in graphene.\cite{WK10,MM10,NG10}
Furthermore, depending on the way how the defects are distributed
over the lattice sites, each kind of disorder can be either
non-correlated or correlated. The non-correlated one corresponds
to the case with uniformly random distributed disorder sources,
i.e., the potential or hopping are randomly changed within a certain
range, or the resonant impurities (vacancies or hydrogen
adatoms) are randomly positioned over the whole lattice; the
correlated one means that the distribution of the disorder follow
particular topological structures, such as Gaussian potentials or
Gaussian hopping parameters, resonant clusters with groups of
vacancies or hydrogen adatoms.
In the present paper, we consider a
noninteracting $\pi$-band tight-binding model on a honeycomb
lattice and solve its time dependent Sch\"odinger equation (TDSE)
to calculate the density of states (DOS). From this, the optical
conductivity is calculated numerically by means of the Kubo formula.

The paper is organized as follows. In Sec. \ref{Sec:Method} we
give details about the method. In Secs. \ref{Sec:NCD} and \ref{Sec:CD} we
show results for the optical conductivity of undoped graphene in the
presence of non-correlated and correlated disorder, respectively. In Sec. %
\ref{Sec:Doped} we calculate the optical spectrum of doped graphene. Our
main conclusions are summarized in Sec. \ref{Sec:Conclusions}.

\section{Model and Method}

\label{Sec:Method}

The tight-binding Hamiltonian of a disordered single-layer graphene is given
by%
\begin{equation}
H=-\sum_{<i,j>}(t_{ij}a_{i}^{\dagger }b_{j}+\mathrm{h.c})+%
\sum_{i}v_{i}c_{i}^{\dagger }c_{i},+H_{imp},  \label{Hamiltonian0}
\end{equation}%
where where $a_{i}^{\dagger }$ ($b_{i}$) creates (annihilates) an electron
on sublattice A (B), $t_{ij}$ is the nearest neighbor hopping parameter, $%
v_{i}$ is the on-site potential, and $H_{imp}$ describes the
hydrogen-like
resonant impurities:%
\begin{equation}
H_{imp}=\varepsilon _{d}\sum_{i}d_{i}^{\dagger}d_{i}+V\sum_{i}\left(
d_{i}^{\dagger}c_{i}+\mathrm{h.c}\right) ,  \label{Eq:Himp}
\end{equation}%
where $\varepsilon _{d}$ is the on-site potential on the
``hydrogen'' impurity (to be specific, we will use this
terminology although it can be more complicated chemical species,
such as various organic groups \cite{WK10}) and $V$ is the
hopping between carbon and hydrogen atoms. For discussions of the
last term see, e.g. Refs.~\onlinecite{Robinson08,WK10,YRK10}. The
spin degree of freedom contributes only through a degeneracy
factor and is omitted for simplicity in Eq.~(\ref{Hamiltonian0}).
A vacancy can be
regarded as an atom (lattice point) with and on-site energy $%
v_{i}\rightarrow \infty $ or with its hopping parameters to other sites
being zero. In the numerical simulation, the simplest way to implement a
vacancy is to remove the atom at the vacancy site.

\begin{figure*}[t]
\begin{center}
\mbox{
\includegraphics[width=7cm]{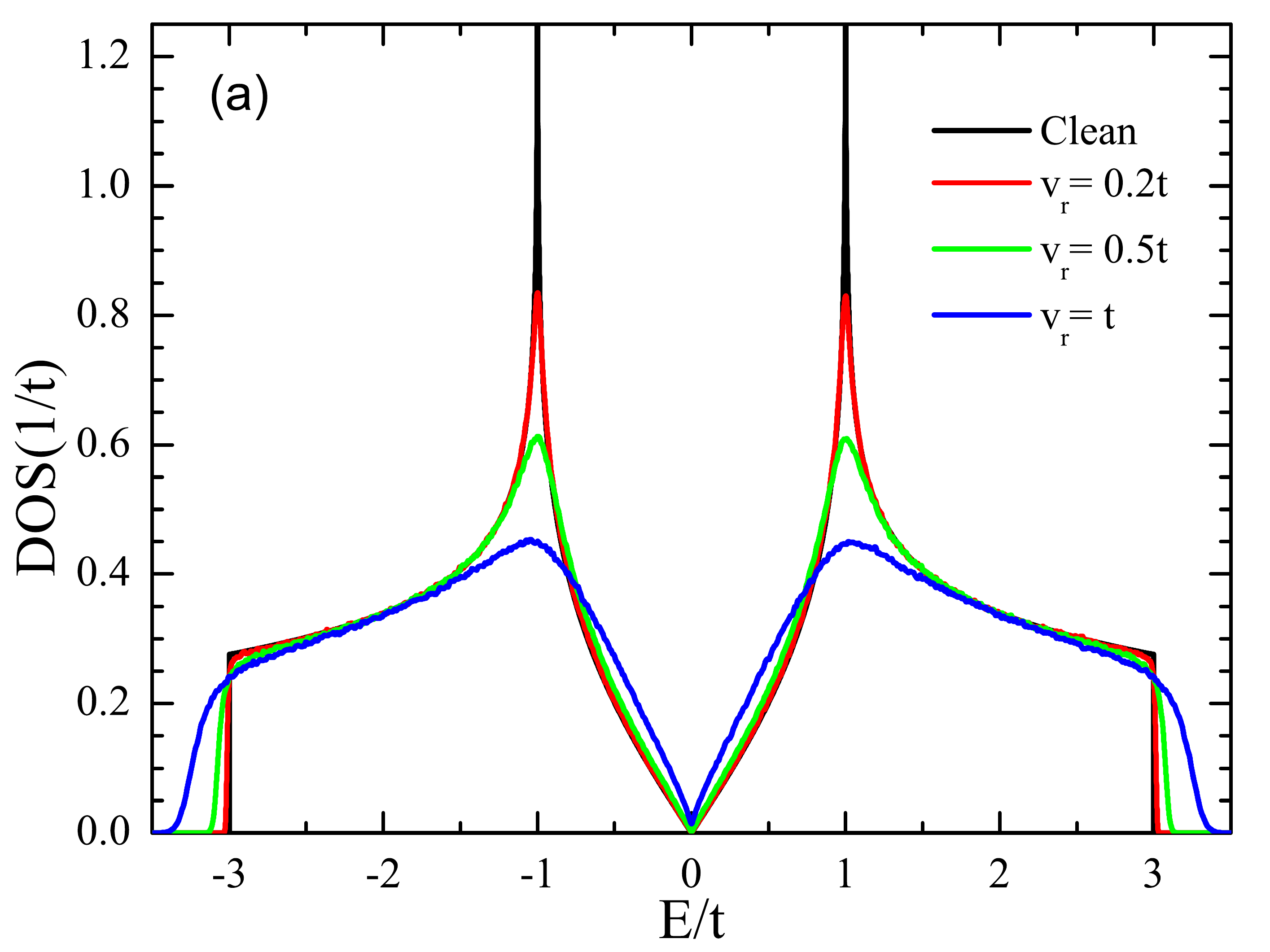}
\includegraphics[width=7cm]{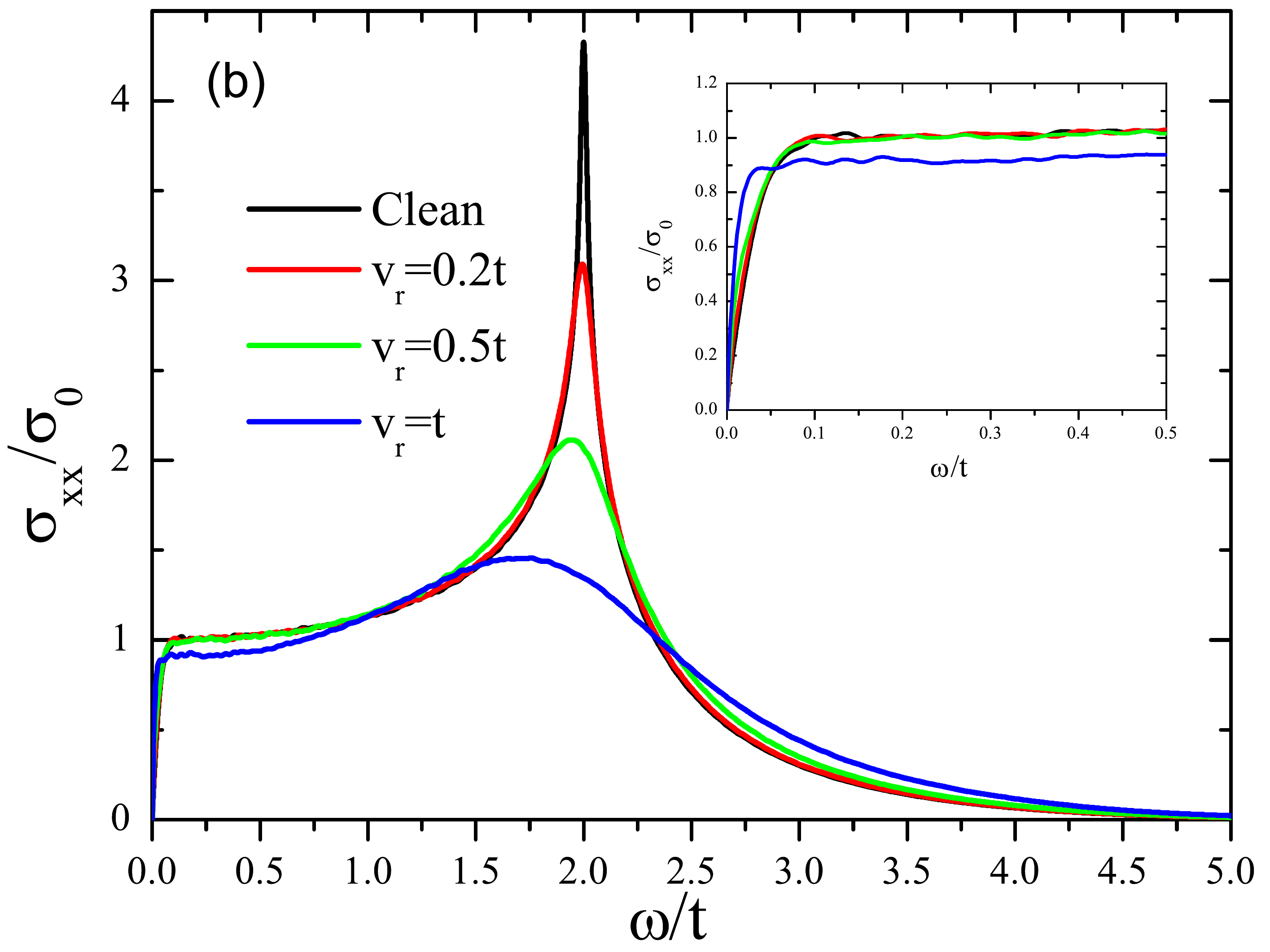}
}
\mbox{
\includegraphics[width=7cm]{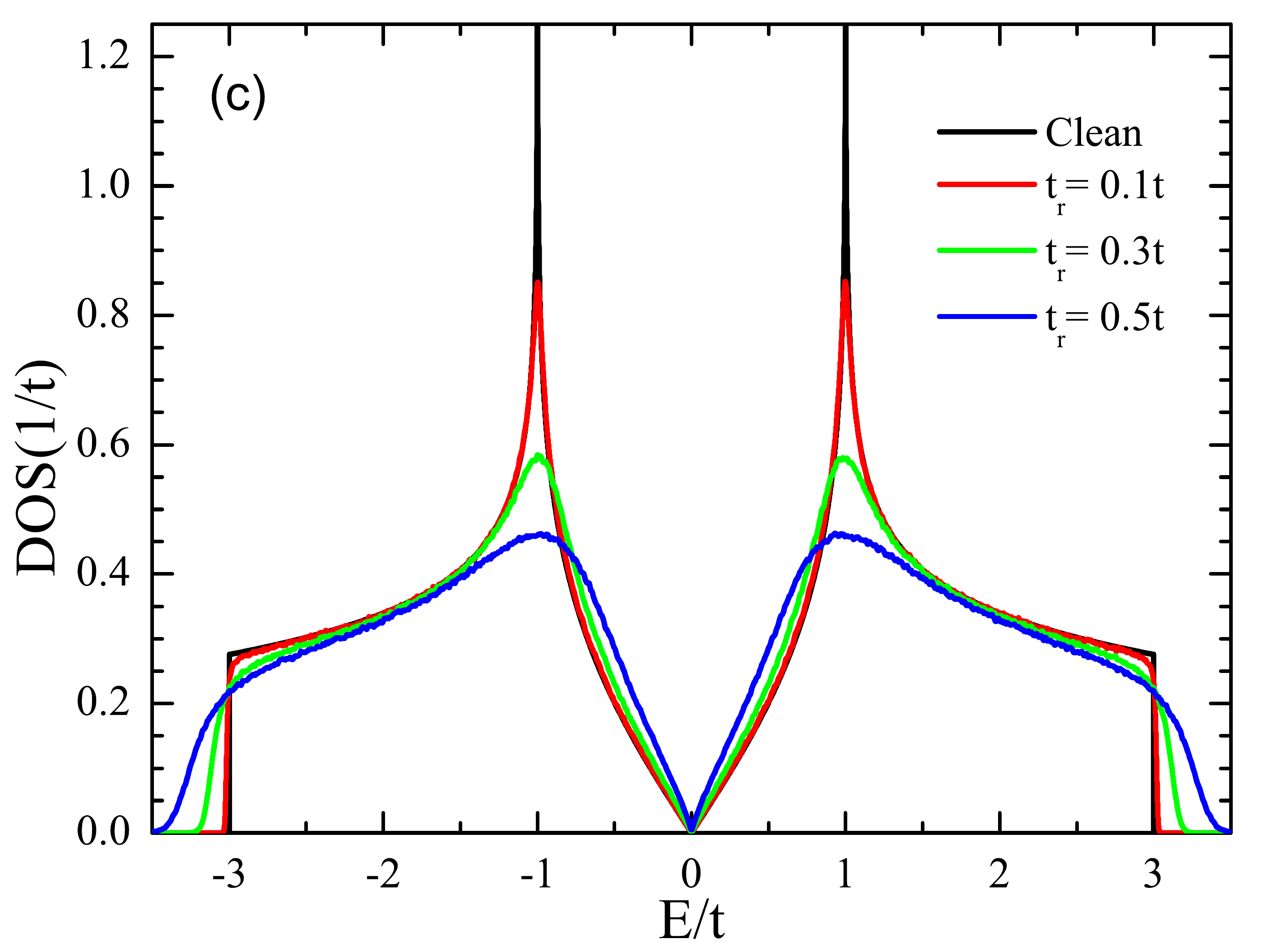}
\includegraphics[width=7cm]{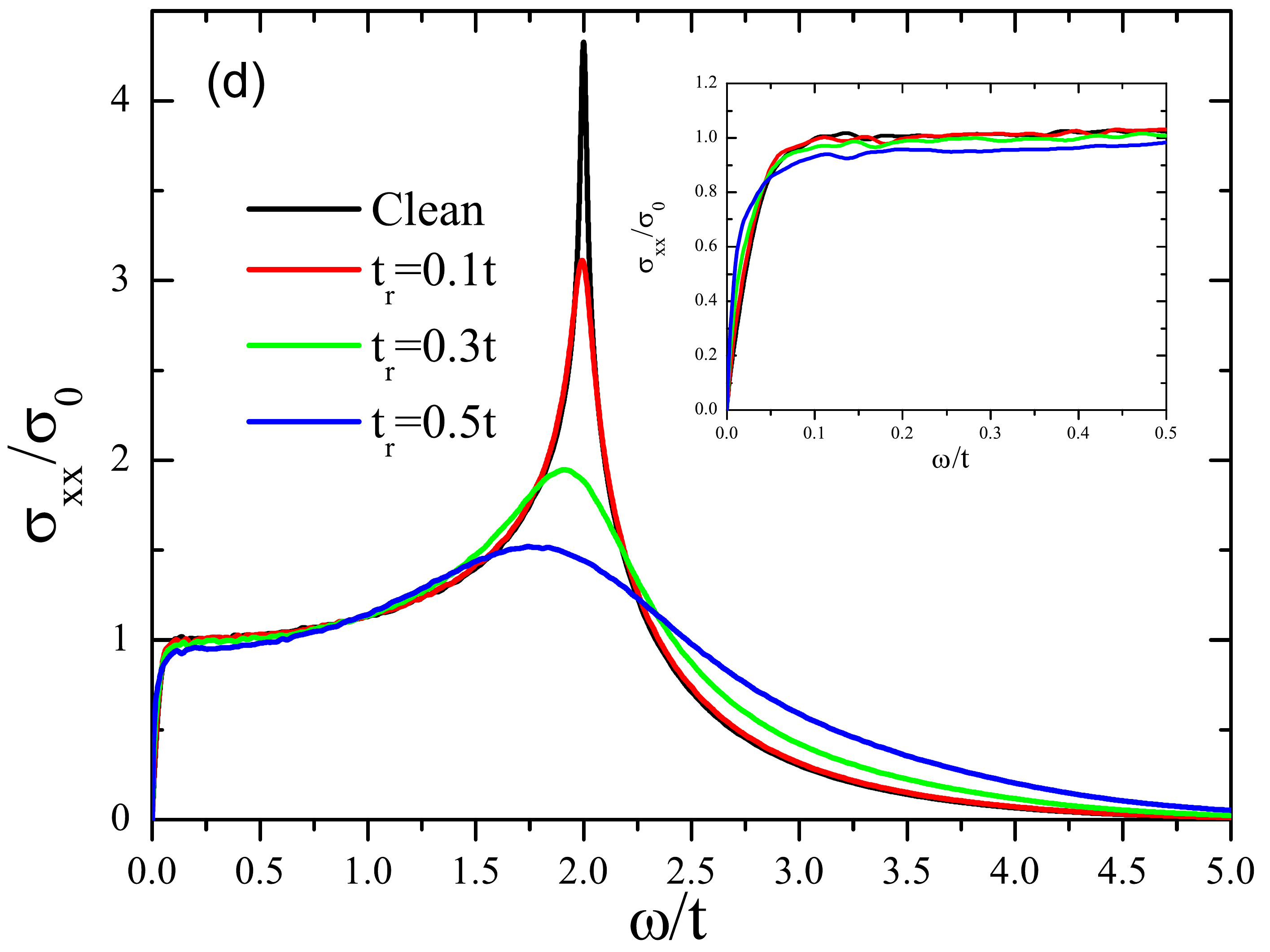}
}
\mbox{
\includegraphics[width=7cm]{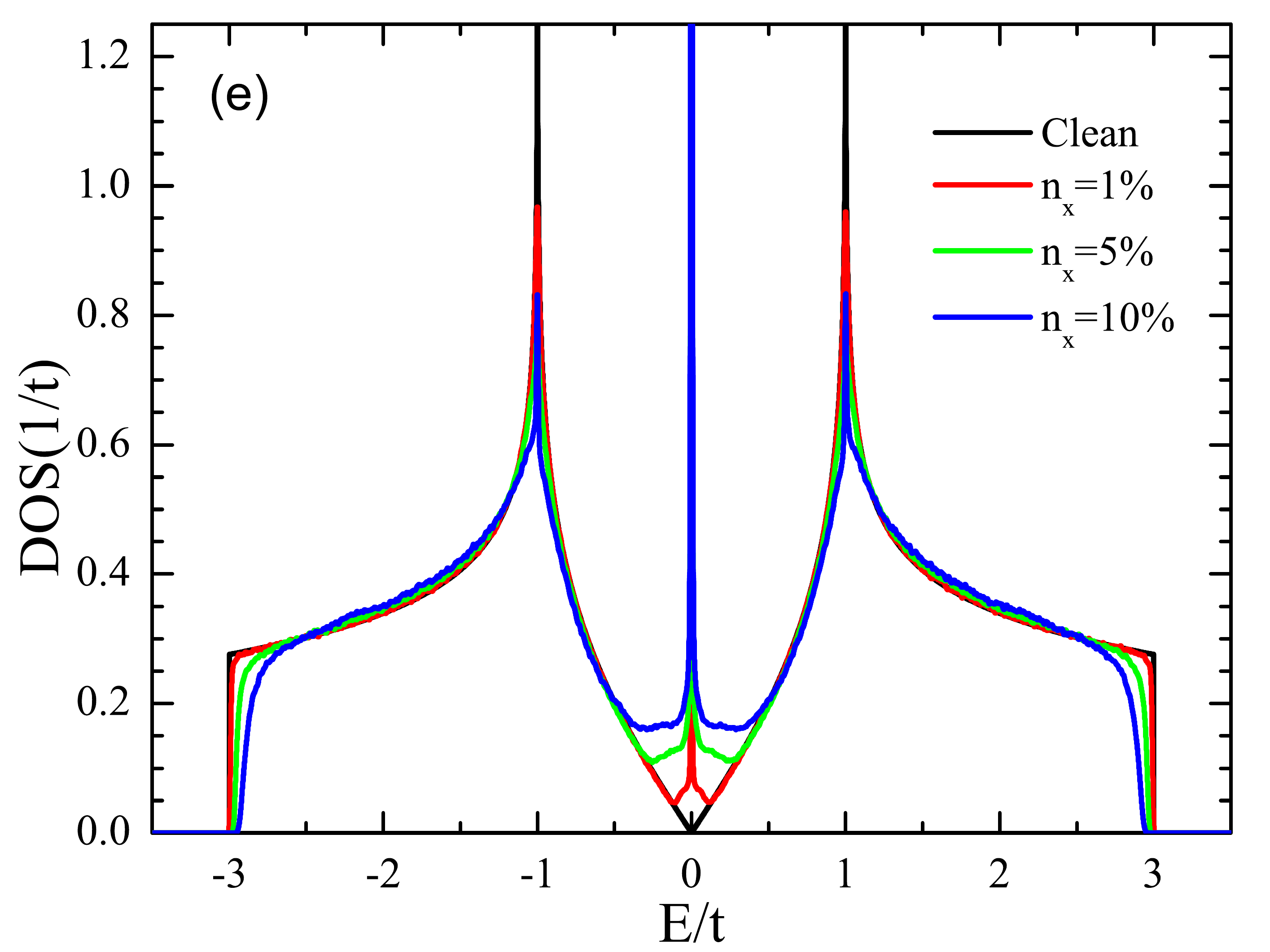}
\includegraphics[width=7cm]{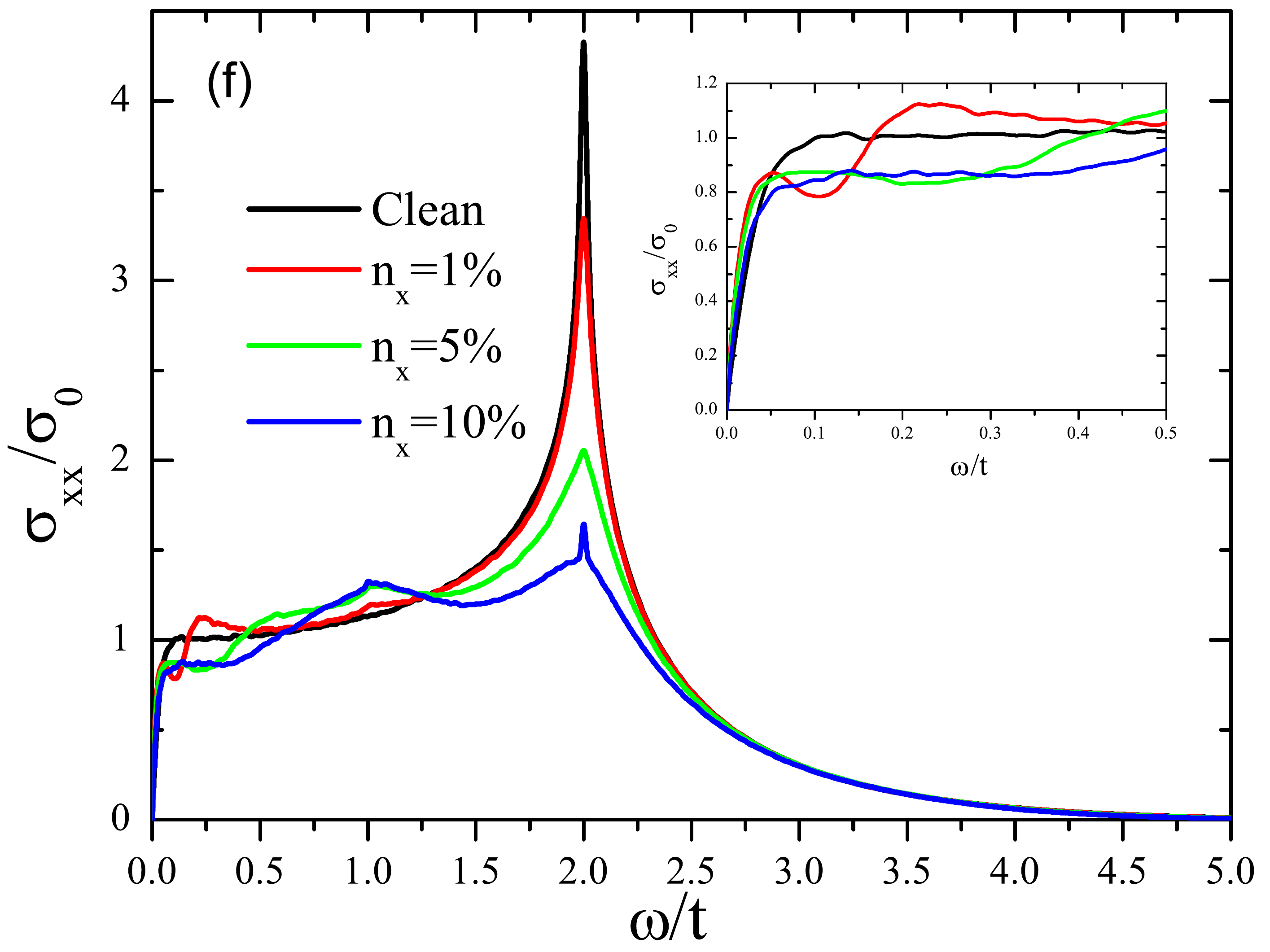}
}
\mbox{
\includegraphics[width=7cm]{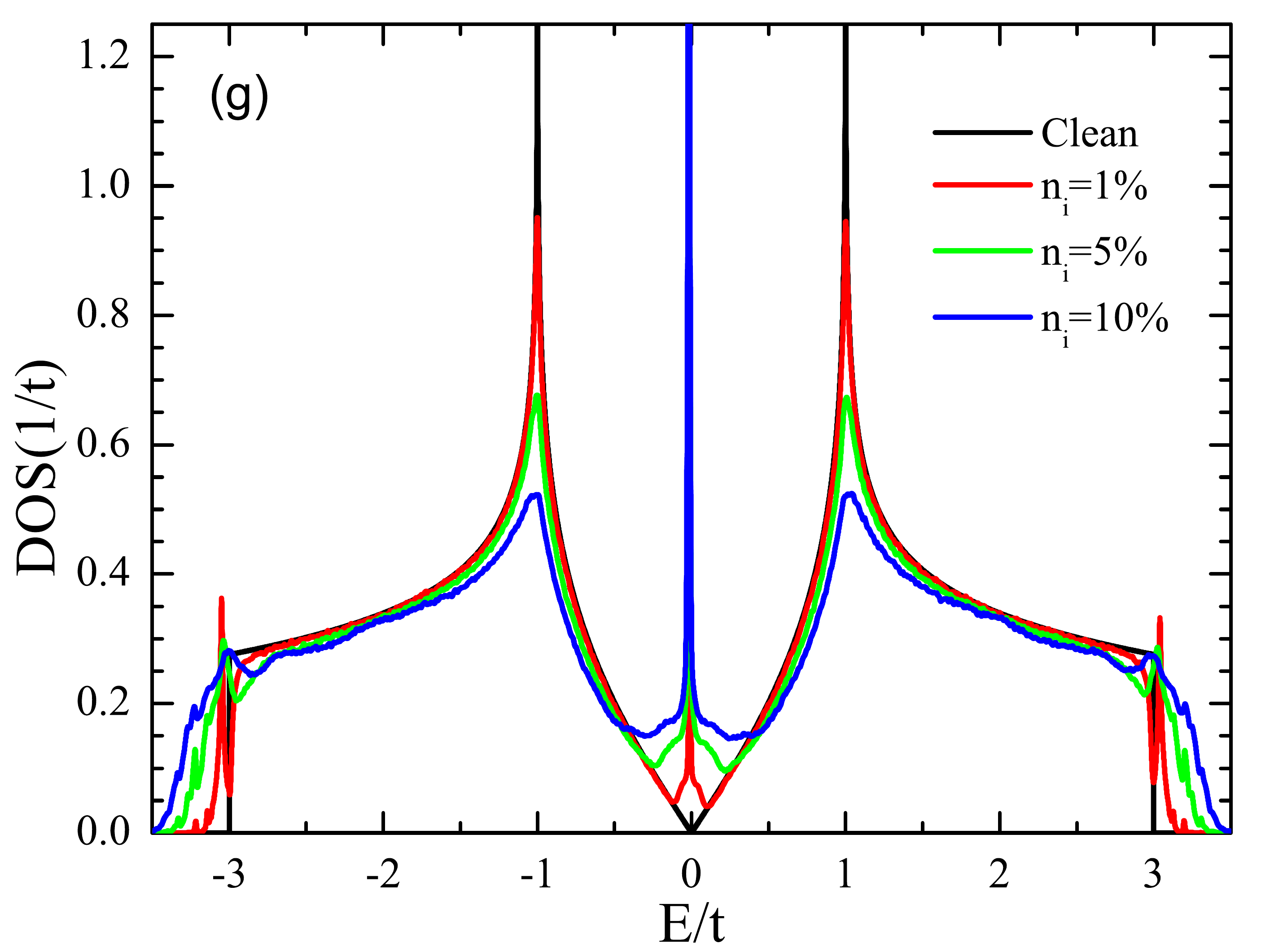}
\includegraphics[width=7cm]{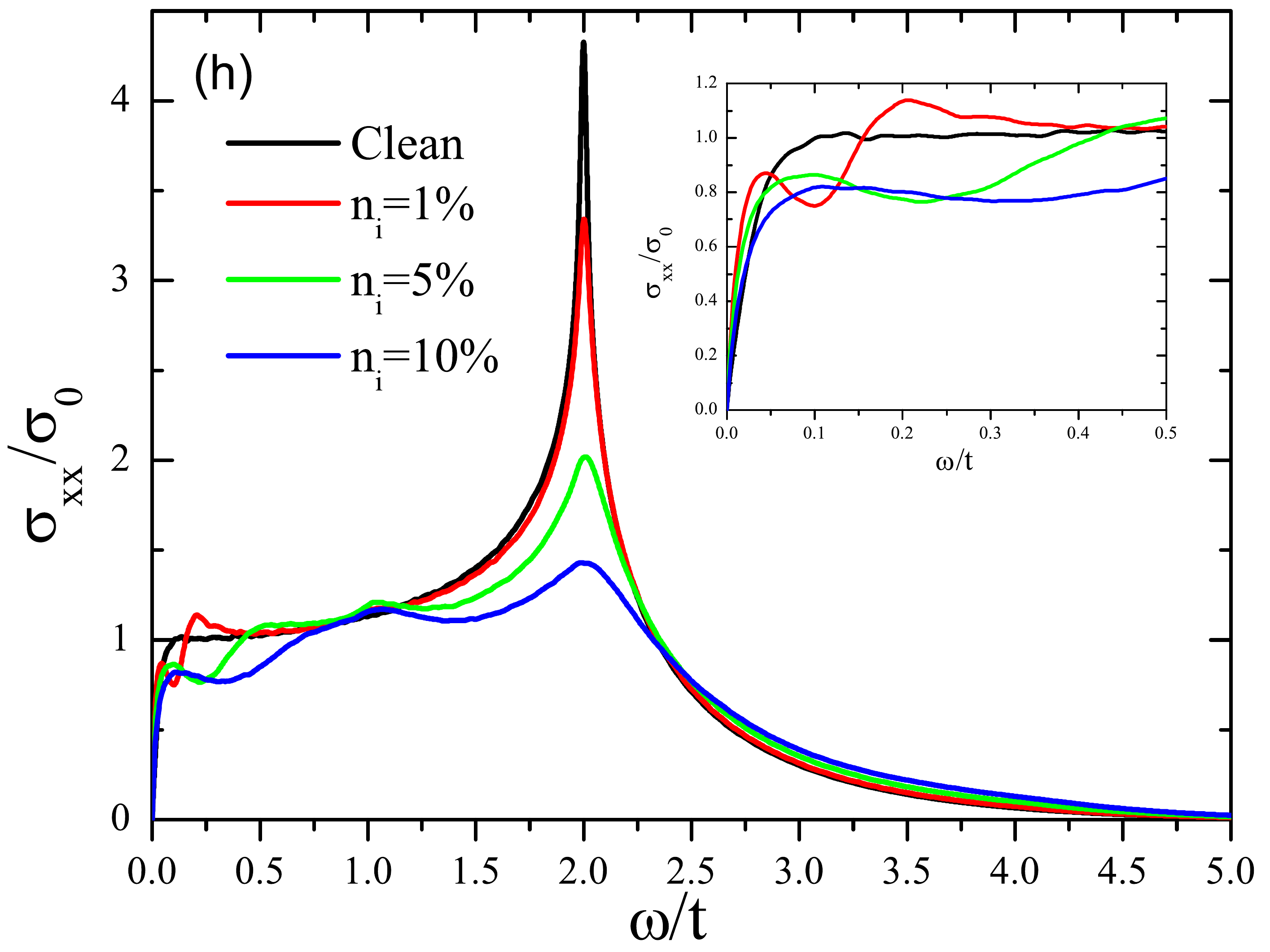}
}
\end{center}
\caption{(Color online) Numerical results for the density of
states (left panels) and optical conductivity (right panels) of
undopped graphene with different kinds of non-correlated
disorders: (a,b) random on-site potentials, (c,d) random hopping
parameters, (e,f) random distributed vacancies, and (g,h) random
distributed hydrogen adatoms. Size of the samples is $4096\times
4096$ for DOS, and $8192\times 8192$ for optical conductivity. In
the right column, the insets show a zoom of the optical
conductivity in the infrared region of the spectrum.}
\label{dos_ac_randomdisorder}
\end{figure*}

\begin{figure*}[t]
\begin{center}
\mbox{
\includegraphics[width=7cm]{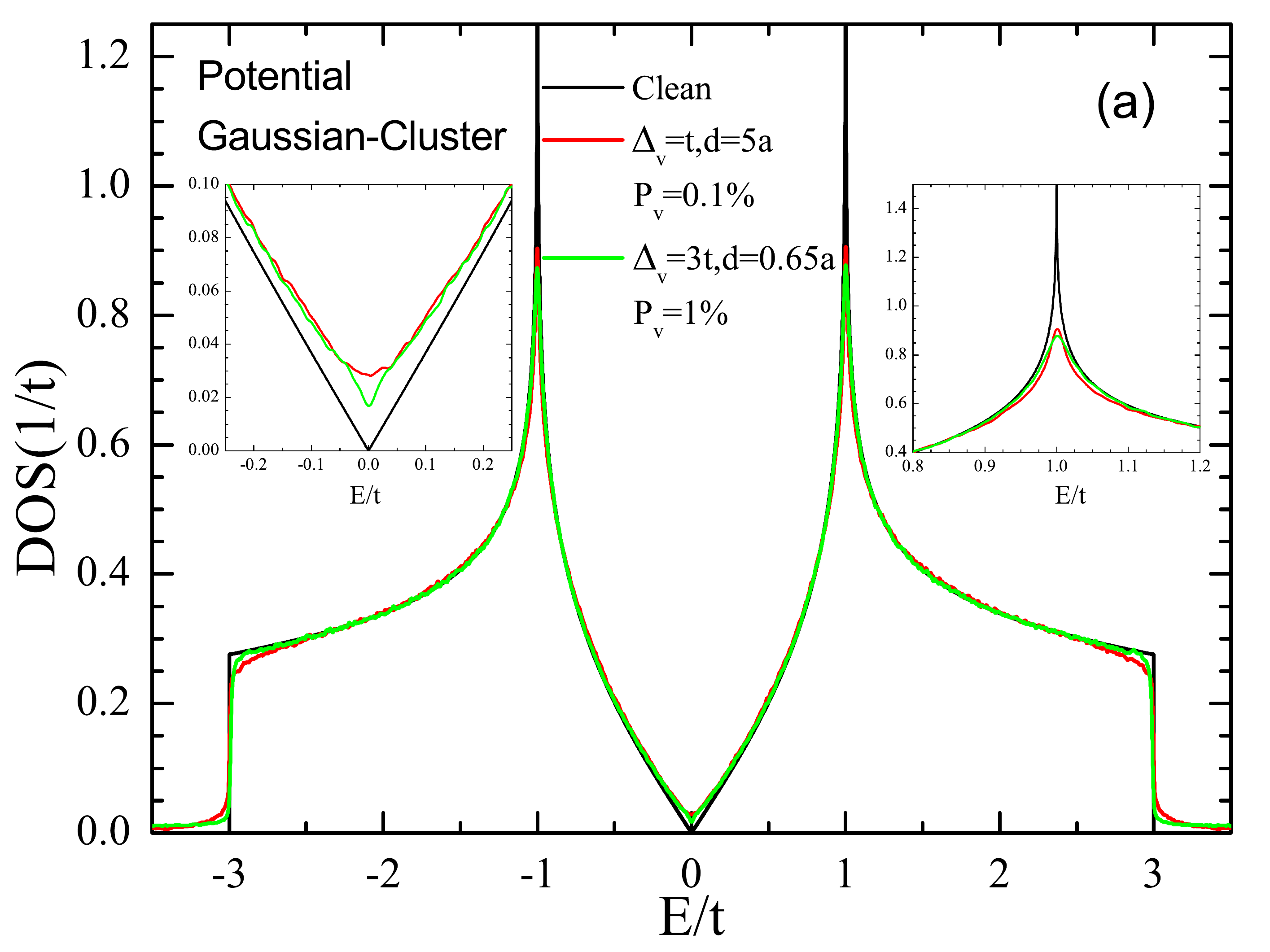}
\includegraphics[width=7cm]{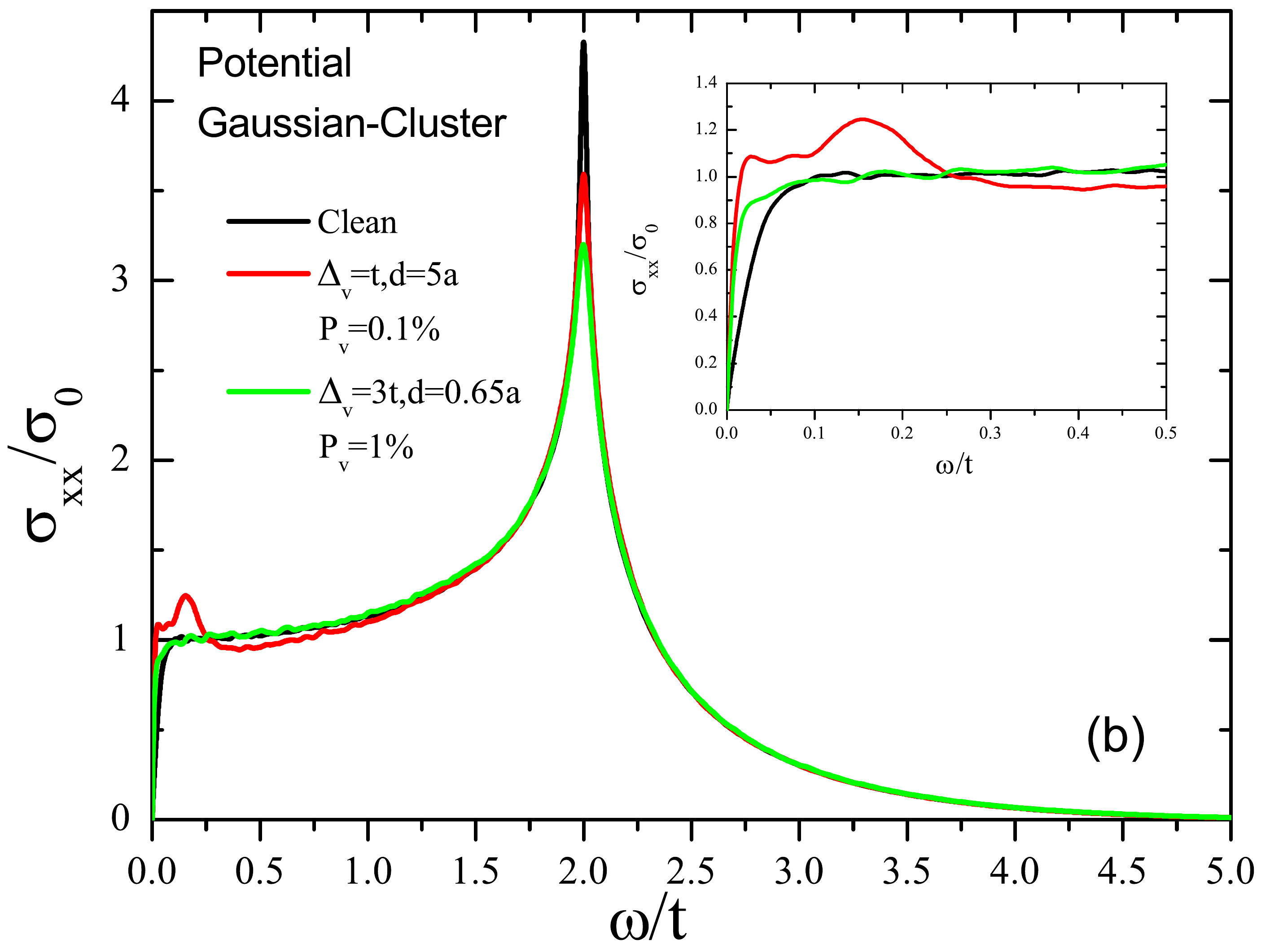}
}
\mbox{
\includegraphics[width=7cm]{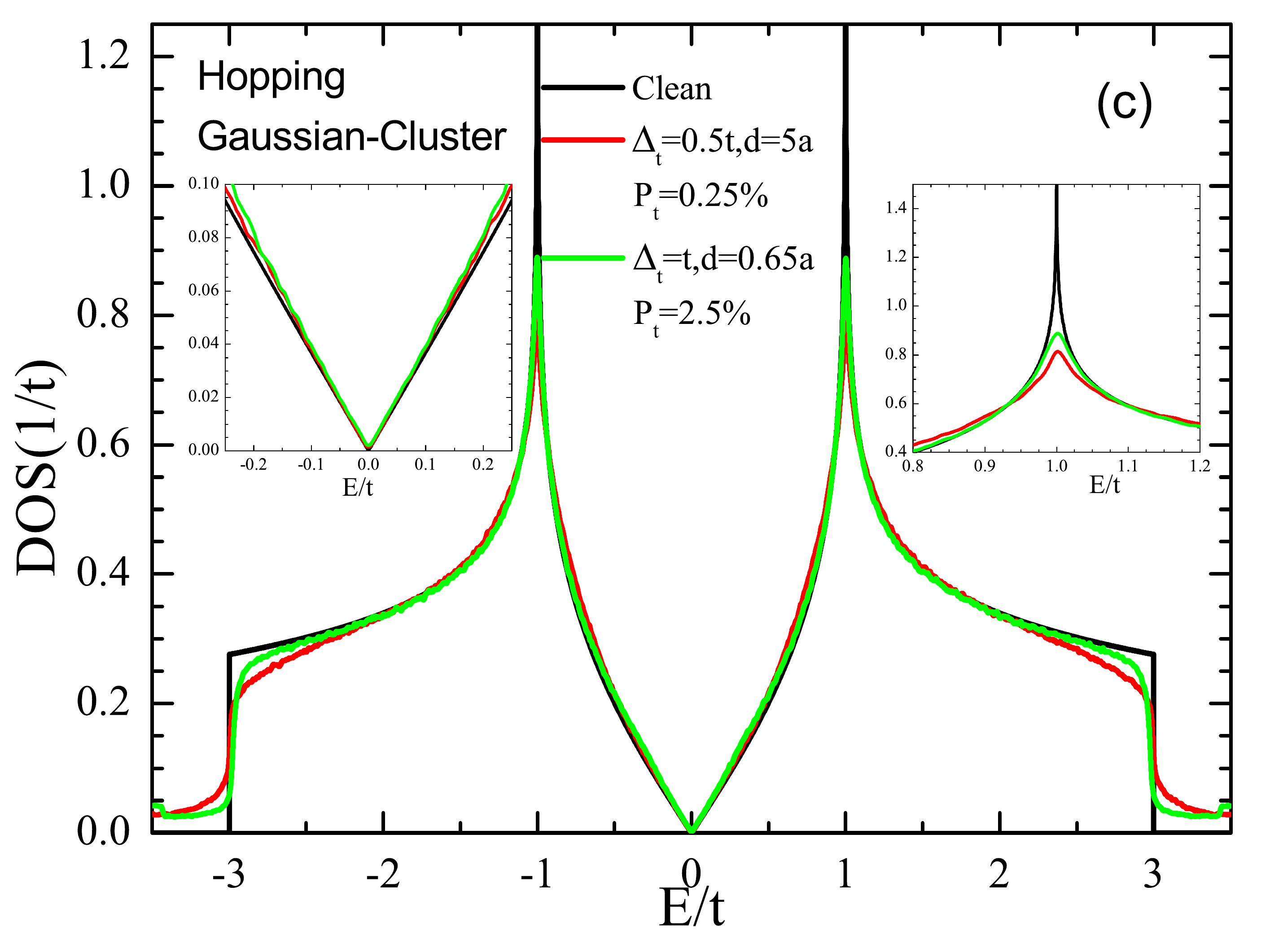}
\includegraphics[width=7cm]{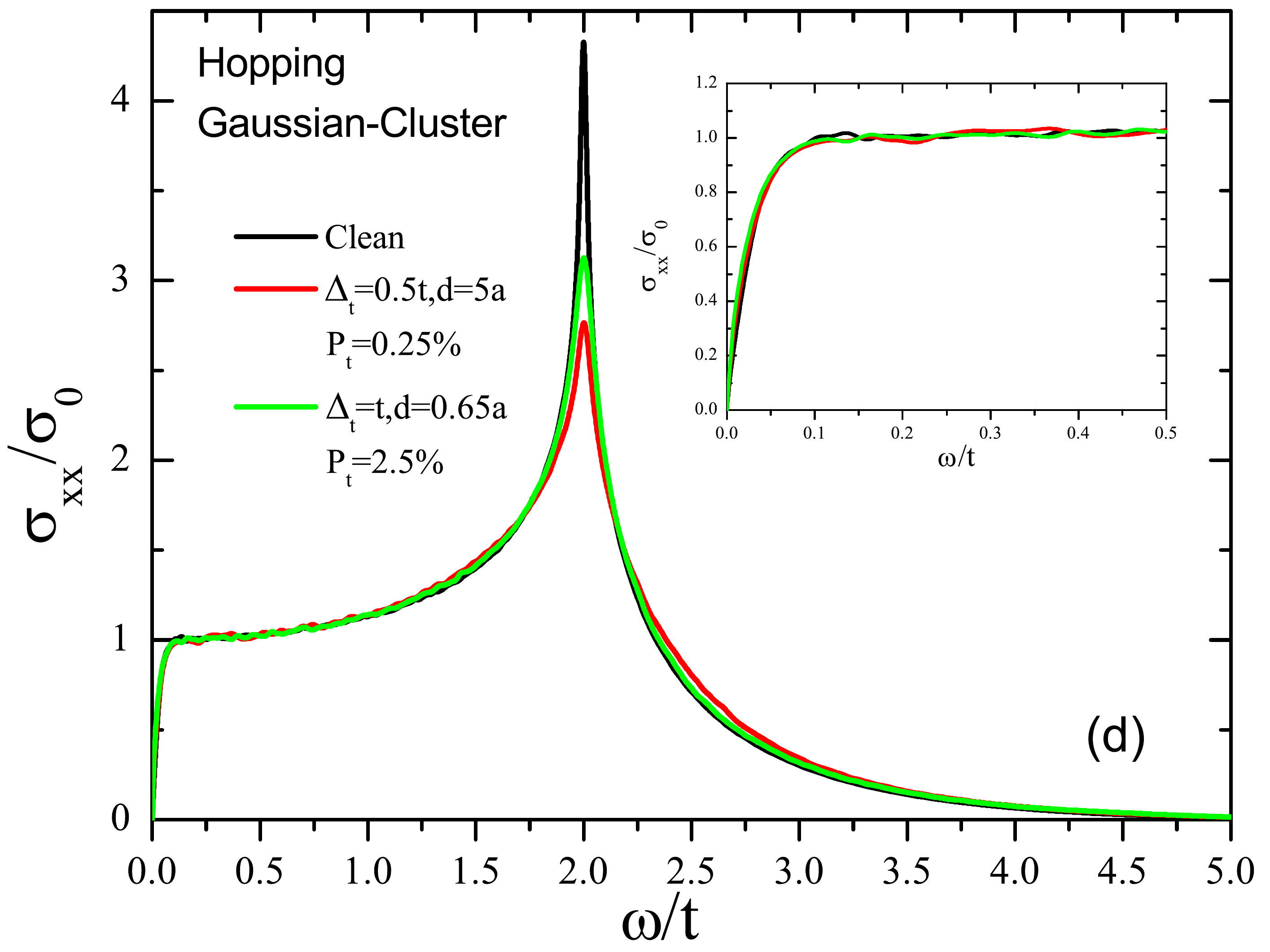}
}
\mbox{
\includegraphics[width=7cm]{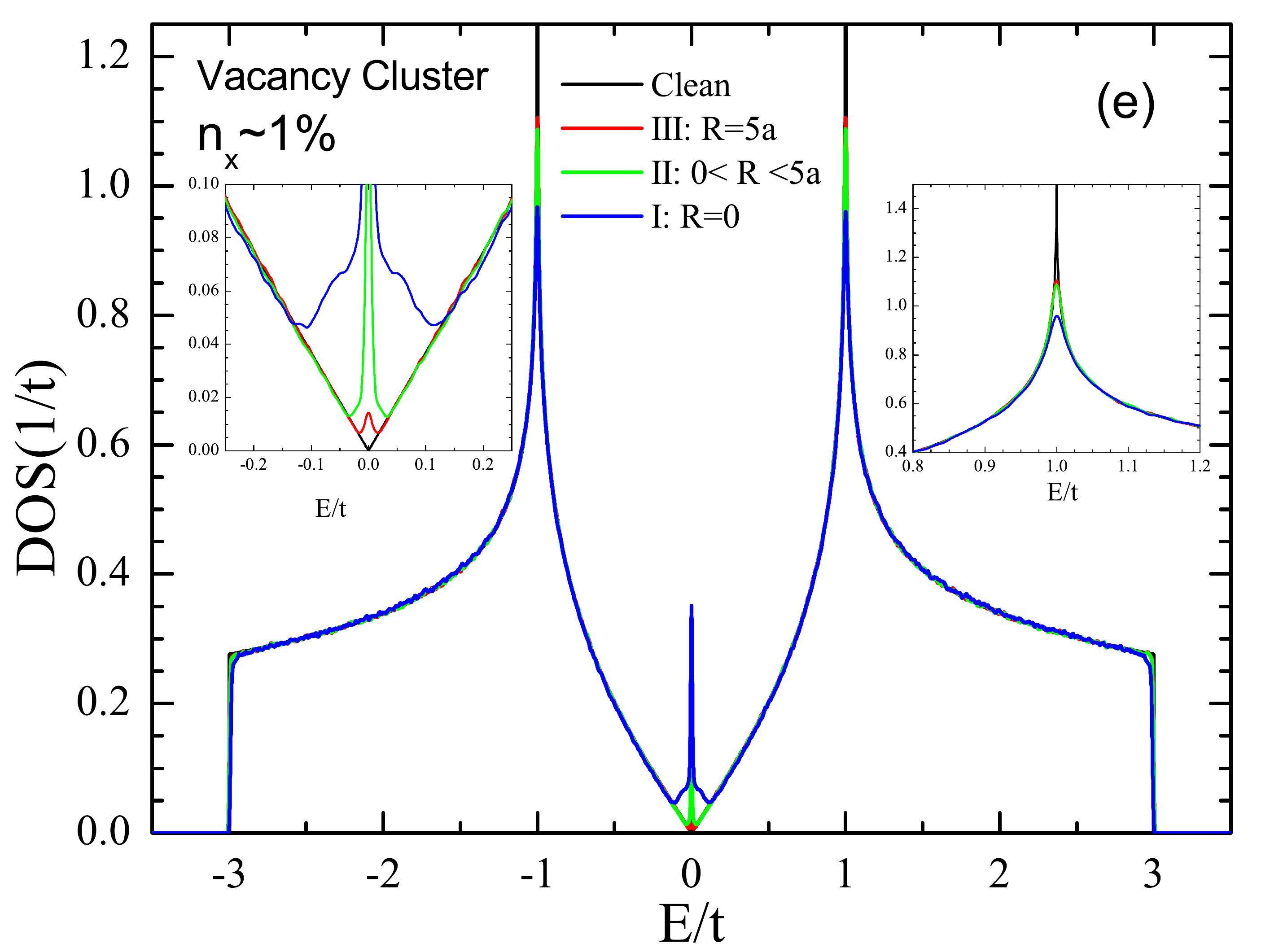}
\includegraphics[width=7cm]{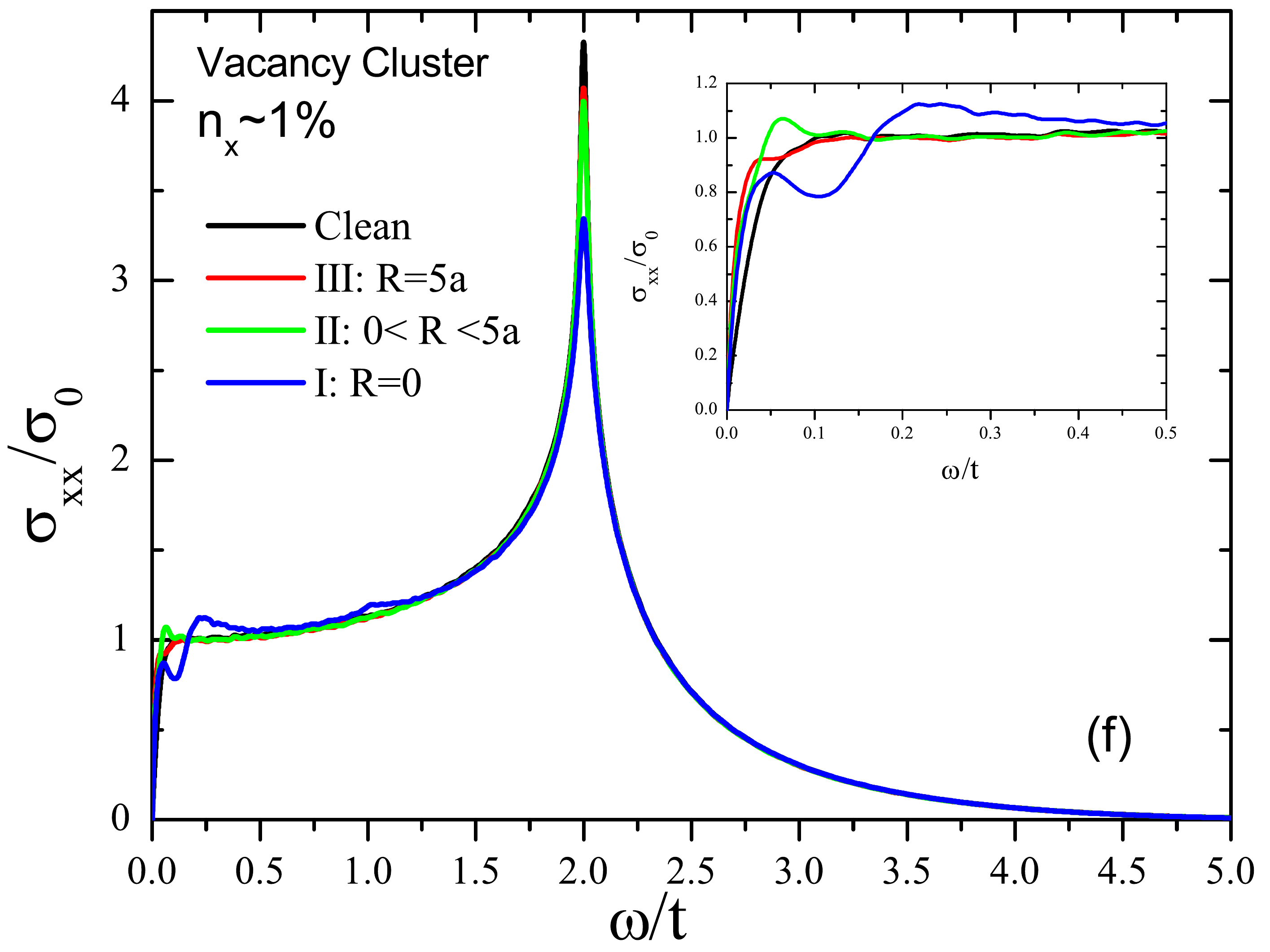}
}
\mbox{
\includegraphics[width=7cm]{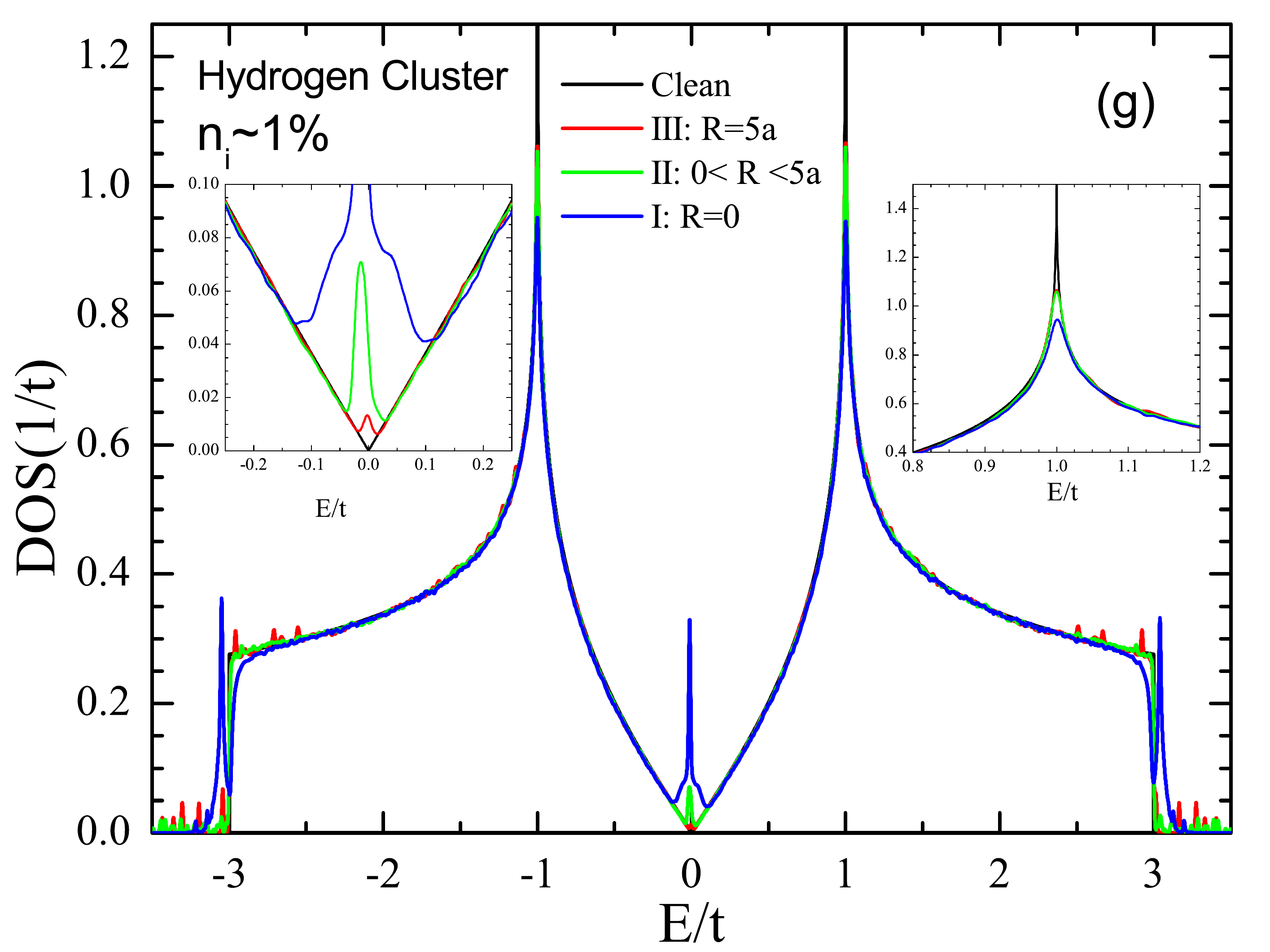}
\includegraphics[width=7cm]{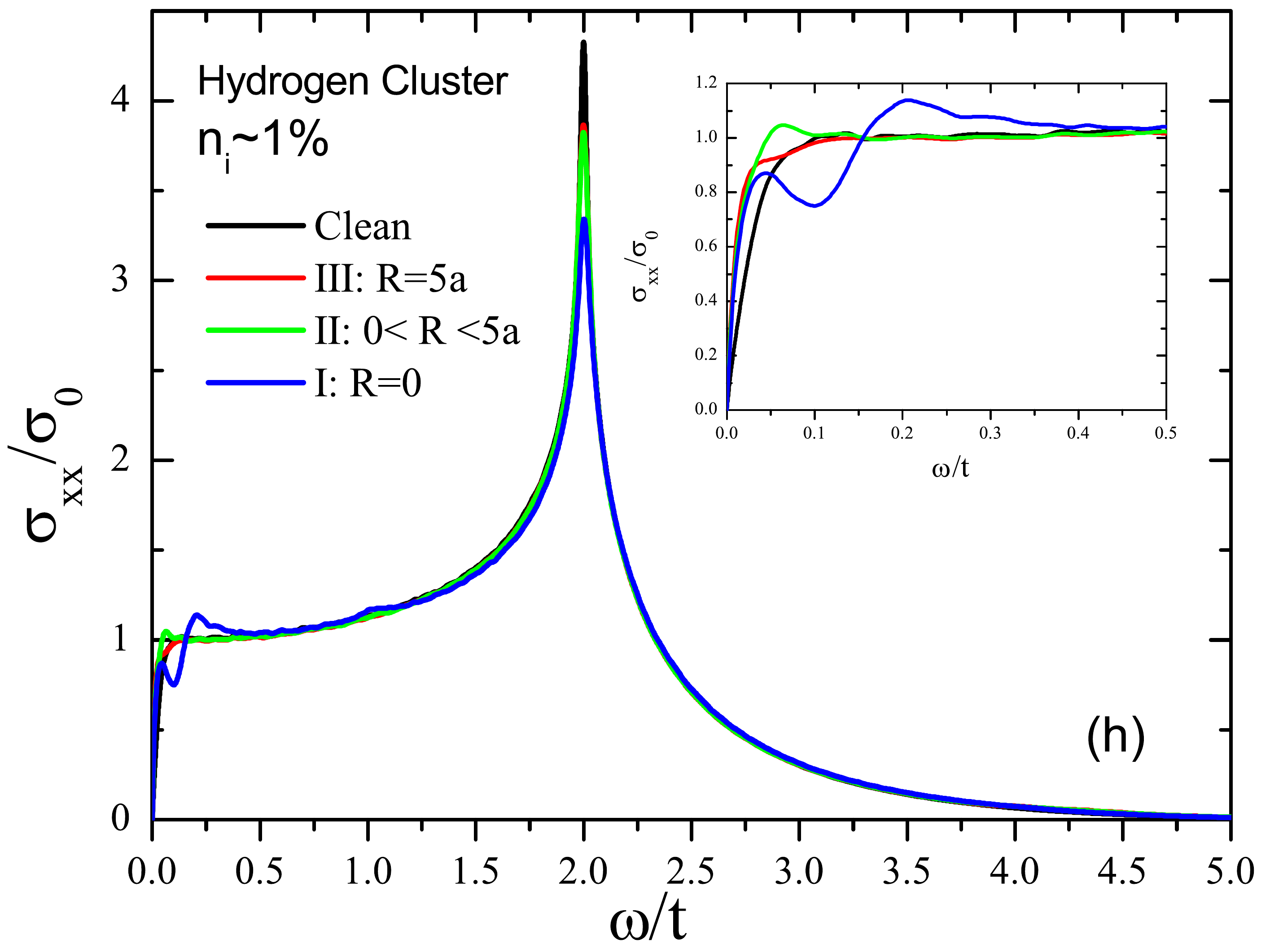}
}
\end{center}
\caption{(Color online) Numerical results for the DOS (left panels) and
optical conductivity (right panels) of undopped graphene with different
kinds of correlated disorders: (a,b) Gaussian potentials, (c,d) Gaussian
hoppings, (e,f) vacancy clusters, and (g,h) hydrogen clusters. The
distribution of the clusters of impurities used for the results (e)-(h) are
sketched in Fig. \protect\ref{figcluster}.}
\label{dos_ac_cluster}
\end{figure*}

The numerical calculations of the optical conductivity and DOS are
performed based on the numerical solution of the TDSE for the
non-interacting particles. In general, the real part of the optical
conductivity contains two parts, the Drude weight $D$ ($\omega =0$) and the
regular part ($\omega \neq 0$). We omit the calculation of the Drude weight,
and focus on the regular part. For
non-interacting electrons, the regular part is \cite{Ishihara1971,YRK10}
\begin{eqnarray}
\sigma _{\alpha \beta }\left( \omega \right)  &=&\lim_{\varepsilon
\rightarrow 0^{+}}\frac{e^{-\beta \omega }-1}{ \omega \Omega }%
\int_{0}^{\infty }e^{-\varepsilon t}\sin \omega t  \notag  \label{gabw2} \\
&&\times 2\text{Im}\left\langle \varphi |f\left( H\right) J_{\alpha }\left(
t\right) \left[ 1-f\left( H\right) \right] J_{\beta }|\varphi \right\rangle
dt,  \notag \\
&&
\end{eqnarray}%
(we put $\hbar =1$) where $\beta =1/k_{B}T$ is the inverse
temperature, $\Omega $ is the sample area, $f\left( H\right)
=1/\left[ e^{\beta \left( H-\mu \right) }+1\right] $ is the
Fermi-Dirac distribution operator, $J_{\alpha }\left( t\right)
=e^{iHt}J_{\alpha }e^{-iHt}$ is the time-dependent current
operator in the $\alpha $ ($=x$ or $y$) direction, and $\left\vert
\varphi \right\rangle $ is a random superposition of all the basis
states in the real space, i.e.,\cite{HR00,YRK10}
\begin{equation}
\left\vert \varphi \right\rangle =\sum_{i}a_{i}c_{i}^{\dagger }\left\vert
0\right\rangle ,  \label{Eq:phi0}
\end{equation}%
where $a_{i}$ are random complex numbers normalized as $\sum_{i}\left\vert
a_{i}\right\vert ^{2}=1$. The time evolution
operator $e^{-iHt}$ and the Fermi-Dirac distribution operator $f\left(
H\right) $ can be obtained by the standard Chebyshev polynomial
representation.\cite{YRK10}

The density of states is calculated by the Fourier transform of the
time-dependent correlation functions \cite{HR00,YRK10}
\begin{equation}
\rho \left( \varepsilon \right) =\frac{1}{2\pi }\int_{-\infty }^{\infty
}e^{i\varepsilon t}\left\langle \varphi \right\vert e^{-iHt}\left\vert
\varphi \right\rangle dt,  \label{Eq:DOS}
\end{equation}%
with the same initial state $\left\vert \varphi \right\rangle $
defined in Eq.~(\ref{Eq:phi0}). For a more detailed description
and discussion of our numerical method we refer to Ref.
\onlinecite{YRK10}. In this paper, we fix the temperature to
$T=300$K. We use periodic boundary conditions in the calculations for
both the optical conductivity and the density of states, and the
size of the system is $8192\times 8192$ or $4096\times 4096$.

\section{Non-Correlated Disorder}

\label{Sec:NCD}

\subsection{Random on-Site Potentials or Nearest-Neighbor Hopping Parameters}

We first consider two different kinds of disorder: random local
change of on-site potentials and random renormalization of the
hopping, which correspond to the diagonal and off-digonal
disorders in the single-layer Hamiltonian
Eq.~(\ref{Hamiltonian0}), respectively. The former acts as a local
shift of the chemical potential of the Dirac fermions, i.e.,
shifts locally the Dirac point, and the latter arises from the
changes of distance or angles between the $p_{z}$ orbitals. In
order to introduce the non-correlated disorders in the on-site
potentials, we consider that the on-site potential $v_{i}$ is
random and uniformly distributed (independently of each site $i$)
between the values $-v_{r}$ and $+v_{r}$. Similarly, the
non-correlated disorder in the nearest-neighbor hopping is
introduced by letting $t_{ij}$ be random and uniformly distributed
(independently of couple of neighboring sites $\langle i,j\rangle$) between
$t-t_{r}$ and $t+t_{r}$. The presence of each type of disorder has
quite similar effect to the density of states [see the
numerical results with different magnitude of disorders in Fig. \ref%
{dos_ac_randomdisorder} (a) and (c) for the random on-site potentials ($%
v_{r}/t=0.2$, $0.5$ and $1$) and random hoppings ($t_{r}/t=0.1$, $0.3$ and $%
0.5$) respectively]. The spectrum is smeared starting from the Van
Hove singularities at $\left\vert E\right\vert =t$, and the
smeared region expands around their vicinal areas as the strength
of the disorder is increased, whereas the spectrum around the
vicinal region of the neutrality point keeps unaffected unless the
disorder is too strong. As the optical conductivity is
proportional to the density of states of the occupied and
unoccupied states, one expects a peak in the spectrum of the
optical conductivity at the energy $\omega \approx 2t$, which
corresponds to
particle-hole excitations between states of the valence band with energy $%
E\approx -t$ and states of the conduction band with energy $E\approx t$.\cite%
{YRK11} These processes contribute to the optical conductivity with a strong
spectral weight due to the enhanced density of states at the Van Hove
singularities of the $\pi $-bands. Because we are considering a
full $\pi$-band tight-binding model for our calculations, this peak is also
present in our results for the optical conductivity, as it is evident in
Figs. \ref{dos_ac_randomdisorder}(b) and (d) at $\omega/t\approx 2$, in
qualitative agreement with recent experimental results.\cite{MSH11} Notice
that the height of the peak is sensitive to the presence of disorder,
getting more and more smeared as the strength of disorder is increased. On
the other hand, for this kind of disorder, for which there is no big change
in the DOS around the Dirac point, one expects that the low energy spectrum
of the optical conductivity should be robust for small disorder, i.e., the
optical conductivity should follow the same spectrum as the clean sample
without any disorder. These expectations are exactly what we observed in the
numerical results of $\sigma \left( \omega \right) $ shown in the insets of
Fig. \ref{dos_ac_randomdisorder} (b) and (d).  This is indeed
the part of the spectrum that can be accounted for within the continuum
(Dirac cone) approximation. We can conclude that the non-correlated random
disorder in the on-site potentials or hopping integrals have almost no
effect on the electronic properties (density of states and AC conductivity)
in the low energy part of the spectrum unless the disorder is too large. On the other hand, the high energy inter-band processes between
states belonging to the Van Hove singularities of the valence and conduction
bands are quite sensitive to the strength of these two kinds of disorder.

\subsection{Random Distributed Vacancies or Hydrogen Impurities}

Next, we consider the influence of two other types of defects on graphene,
namely, vacancies and hydrogen impurities. Introducing vacancies in a
graphene sheet will create a zero energy mode (midgap state), effect that has been anticipated in many theoretical works,\cite%
{Peres2006,Pereira2006,Pereira2008,YRK10} and which has been recently observed experimentally by means of scanning tunneling spectroscopy (STM) measurements.\cite{Ugeda10}
It is shown that the number of
midgap states increases with the concentration of the vacancies \cite{YRK10}%
, and the inclusion of vacancies brings an increase of spectral weight to
the surrounding of the Dirac point ($E=0)$ and smears the van Hove
singularities.\cite{Peres2006,Pereira2008,YRK10} This is in fact the
behavior found in Fig. \ref{dos_ac_randomdisorder} (e) for the DOS of
graphene with different concentrations of vacancies $n_{x}$, where the
numerical results with $n_{x}=1\%$, $5\%$, $10\%$ are represented and
compared to the density of states of clean graphene.

The presence of hydrogen impurities, which are introduced by the formation
of a chemical bond between a carbon atom from the graphene sheet and a
carbon/oxygen/hydrogen atom from an adsorbed organic molecule (CH$_{3}$, C$%
_{2}$H$_{5}$, CH$_{2}$OH, as well as H and OH groups) have quite similar
effect to the electronic structure and transport properties of graphene.\cite%
{WK10,YRK10} The adsorbates are described by the Hamiltonian $H_{imp}$ in
Eq.~(\ref{Hamiltonian0}). The band parameters $V\approx 2t$ and $\epsilon
_{d}\approx -t/16$ are obtained from the \textit{ab initio} density
functional theory (DFT) calculations.\cite{WK10} Following Refs. %
\onlinecite{WK10,YRK10}, we call these impurities as adsorbates hydrogen
atoms but actually, the parameters for organic groups are almost the same.
\cite{WK10} As we can see from Fig. \ref{dos_ac_randomdisorder} (g), small
concentrations of hydrogen impurities have similar effects as the same
concentration of vacancies to the density of states of graphene. Hydrogen
adatoms also lead to zero modes and the quasilocalization of the low-energy
eigenstates, as well as to a smearing of the Van Hove singularities. The
shift of the central peak of the density of states with respect to the Dirac
point in the case of hydrogen impurities is due to the nonzero (negative)
on-site potentials $\epsilon _{d}$.

The similarity in the density of states leads to similar optical spectra for
graphene with random vacancies or hydrogen adatoms, as it can be seen in
Fig. \ref{dos_ac_randomdisorder} (f) and (h). In the high and
intermediate energy part of the spectrum it is noticeable, apart from the
smearing of the $\omega\approx 2t$ peak due to the renormalization of the
Van Hove singularities, the appearance of a new peak at an energy $%
\omega\approx t$. This peak is associated to optical transitions between the
newly formed midgap states (with energy $E\approx 0$) and the states of the
Van Hove singularities (with energy $E\approx t$). Notice that, contrary to
the $\omega\approx 2t$ peak, the height of this $\omega\approx t$ peak grows
with the strength of disorder, due to the enhancement of the DOS at the
Dirac point. Therefore, we expect that this peak should be observed in
optical spectroscopy measurements of graphene samples with sufficient amount
of resonant scatterers.

In the low energy part of the spectra, the new structure of
the DOS around the Dirac point leads to a modulation of the infrared
conductivity, as it can be seen in the insets of Figs. \ref%
{dos_ac_randomdisorder} (f) and (h). The lower peaks, which in Figs. \ref%
{dos_ac_randomdisorder} (f) and (h) corresponds to a conductivity
$\sigma \approx 0.9\sigma _{0}\,$ for different concentration of
impurities, might have their origin from excitations involving
states surrounding the zero modes (central high peak in the
density of states). At slightly higher energies there is a new set
of peaks that can be associated to processes involving states at
the boundaries of the midgap states. The optical conductivities in
the region between these two peaks are in general smaller compared
to those in clean graphene, which can be due to the fact that the
midgap states are quasilocalized states.

\section{Correlated Disorders}

\label{Sec:CD}

\subsection{Gaussian Potentials and Gaussian Hoppings}

As discussed in the previous section, the change of on-site
potential can be regarded as a local chemical potential shift for the Dirac
fermions. If the random potentials are too large, characteristics of the
graphene band structure such as the Dirac points or the Van Hove
singularities can disappear completely, and the whole spectrum becomes
relatively flat over the whole energy range.\cite{YRK10b} Therefore in order to introduce large
values of random potentials but keep a relatively similar spectrum, in this
section we use small concentrations of correlated Gaussian potentials,
defined as \cite{LMC08,YRK10b}
\begin{equation}
v_{i}=\sum_{k=1}^{N_{imp}^{v}}U_{k}\exp \left( -\frac{\left\vert \mathbf{r}%
_{i}-\mathbf{r}_{k}\right\vert ^{2}}{2d^{2}}\right) ,  \label{vgaussian}
\end{equation}%
where $N_{imp}^{v}$ is the number of the Gaussian centers, which are chosen
to be randomly distributed over the carbon atoms ($\mathbf{r}_{k}$), $U_{k}$ is
uniformly random in the range $[-\Delta _{v},\Delta _{v}]$ and $d$ is
interpreted as the effective potential radius. The typical values of $d$
used in our model are $d=0.65a$ and $5a$ for short- and long-range Gaussian
potential, respectively. Here $a\approx 1.42$\r{A}~ is the carbon-carbon
distance in the single-layer graphene. The value of $N_{imp}^{v}$ is
characterized by the ratio $P_{v}=N_{imp}^{v}/N$, where $N$ is the total
number of carbon atoms of the sample. As one can see from Fig. \ref%
{dos_ac_cluster}(a), in the presence of locally strong disorders ($\Delta
_{v}=3t$ and $t$ for short and long range Gaussian potentials, respectively)
the whole spectrum of DOS is quite similar to the case of clean graphene,
but with the emergence of states in the vicinal area around the Dirac point,
and also a smearing of the Van Hove singularities. This kind of
disorder leads to regions of the graphene membrane where the Dirac point is
locally shifted to the electron ($U_{k}<0$) or to the hole ($U_{k}>0$) side
with the same probability, rising the DOS at zero energy. The final spectrum
is similar to the one of clean graphene but with a series of electron-hole
puddles which are formed at the maxima and minima of the potential.  The enhancement of the DOS around the Dirac point leads to the
possibility for new excitations in the low energy part spectrum, as compared
to the clean case, as it can see in Fig. \ref{dos_ac_cluster}(b). For
the cases we consider, the presence of long-range Gaussian potentials change the low energy optical
spectrum completely with the emergence of a new peak around $\omega \approx
0.15t$. The optical conductivity in the region $\omega <0.24t$ is larger
than in clean graphene but becomes smaller for $\omega >0.24t$. The increase of
the conductivity might have its origin in the possible excitations between
electron and hole puddles. Indeed, the renormalization of the spectrum
obtained by considering long-range Gaussian potentials leads to a larger
optical contribution than for short-range Gaussian potentials, which
yield infra-red spectra much more close to that of a clean graphene
membrane.

The local strong disorder in the hopping between carbon atoms is introduced
in a similar way as the correlated potentials, i.e., with a distribution of
the nearest-neighbor hopping parameter given by\cite{YRK10b}%
\begin{equation}
t_{ij}=t+\sum_{k=1}^{N_{imp}^{t}}T_{k}\exp \left( -\frac{\left\vert \mathbf{r%
}_{i}+\mathbf{r}_{j}-2\mathbf{r}_{k}\right\vert ^{2}}{8d_{t}^{2}}\right) ,
\label{tgaussian}
\end{equation}%
where $N_{imp}^{t}$ is the number of the Gaussian centers ($\mathbf{r}_{k}$%
), $T_{k}$ is uniformly random in the range $[-\Delta _{t},\Delta _{t}]$ and
$d_{t}$ is interpreted as the effective screening length. Similarly, the
typical values of $d_{t}$ are the same as for the Gaussian potential, i.e., $%
d_{t}=0.65a$ and $5a$ for short- and long-range Gaussian random hopping,
respectively, and the values of $N_{imp}^{t}$ are characterized by the ratio
$P_{t}=N_{imp}^{t}/N$.

Numerical results for the DOS and optical conductivity of graphene with
short- ($\Delta _{t}=3t,$ $d_{t}=0.65a$) and long-range ($\Delta _{t}=1t,$ $%
d_{t}=5a$) Gaussian hoppings are shown in Fig. \ref{dos_ac_cluster}(c-d). This kind of disorder accounts for the effect of substitutional
impurities like B or N instead of C, or local distortions of the membrane.
Concerning the physics around the neutrality point, in this case the Dirac
point remains unchanged although there is a local renormalization of the
slope of the band. As a consequence, the Fermi velocity around Dirac is
locally increased (when $t_{k}>0$) or decreased (when $t_{k}<0$). However,
no midgap states are created by this kind of disorder, and the DOS remains
quite similar to the one of a clean graphene layer, as it can be seen in
Fig. \ref{dos_ac_cluster}(c). In particular, the absence of an impurity band
at $E\approx 0$ makes that the optical conductivity presents only slight
deviations as compared to the clean case. This can be seen in Fig. \ref%
{dos_ac_cluster}(d), where (apart from the smearing of the Van Hove peak)
the optical spectrum, especially in the infra-red region, remains
practically the same as in the absence of disorder.

\subsection{Vacancy Clusters and Hydrogen Clusters}

\begin{figure}[t]
\begin{center}
\mbox{
\includegraphics[width=4cm]{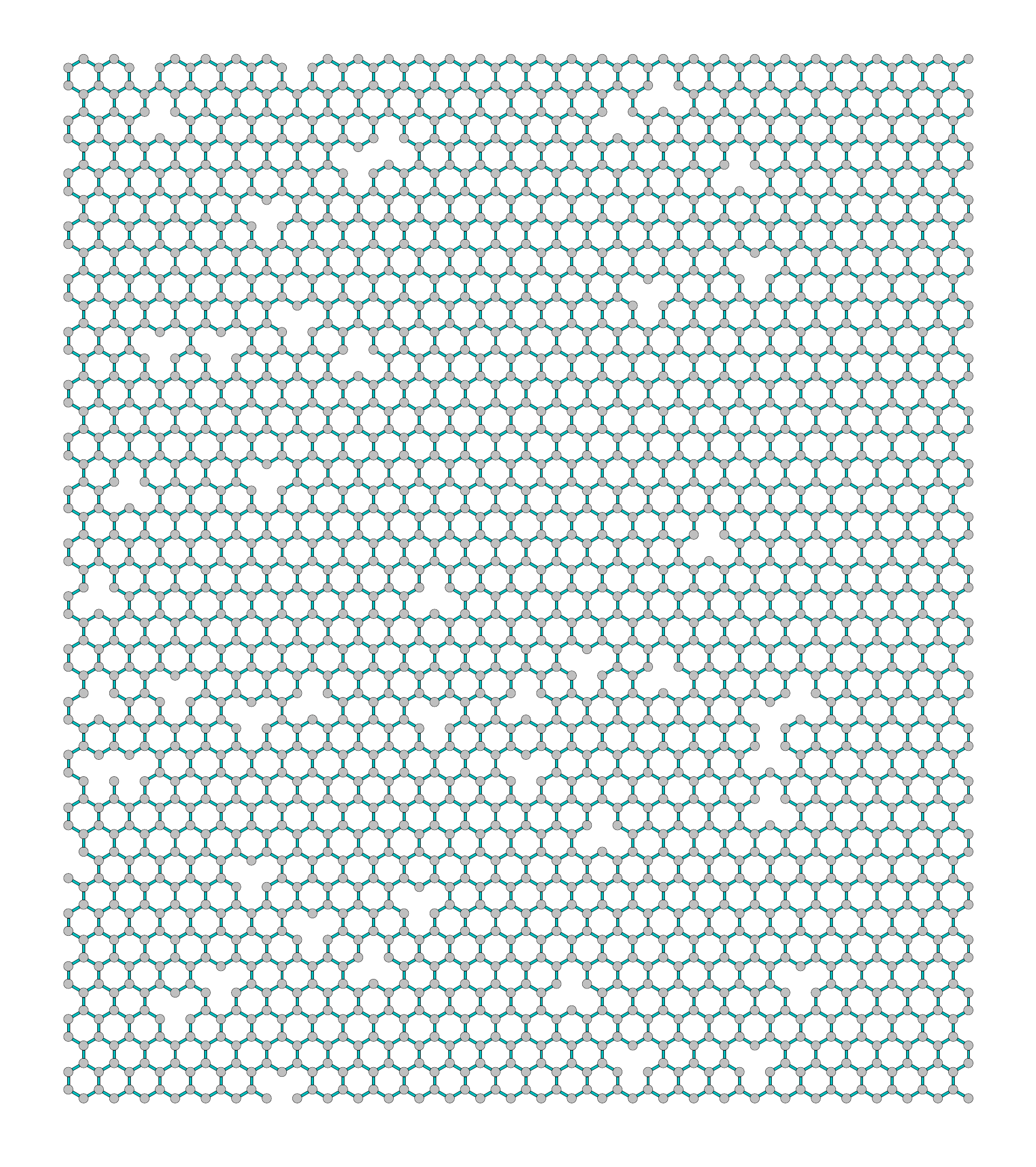}
\includegraphics[width=4cm]{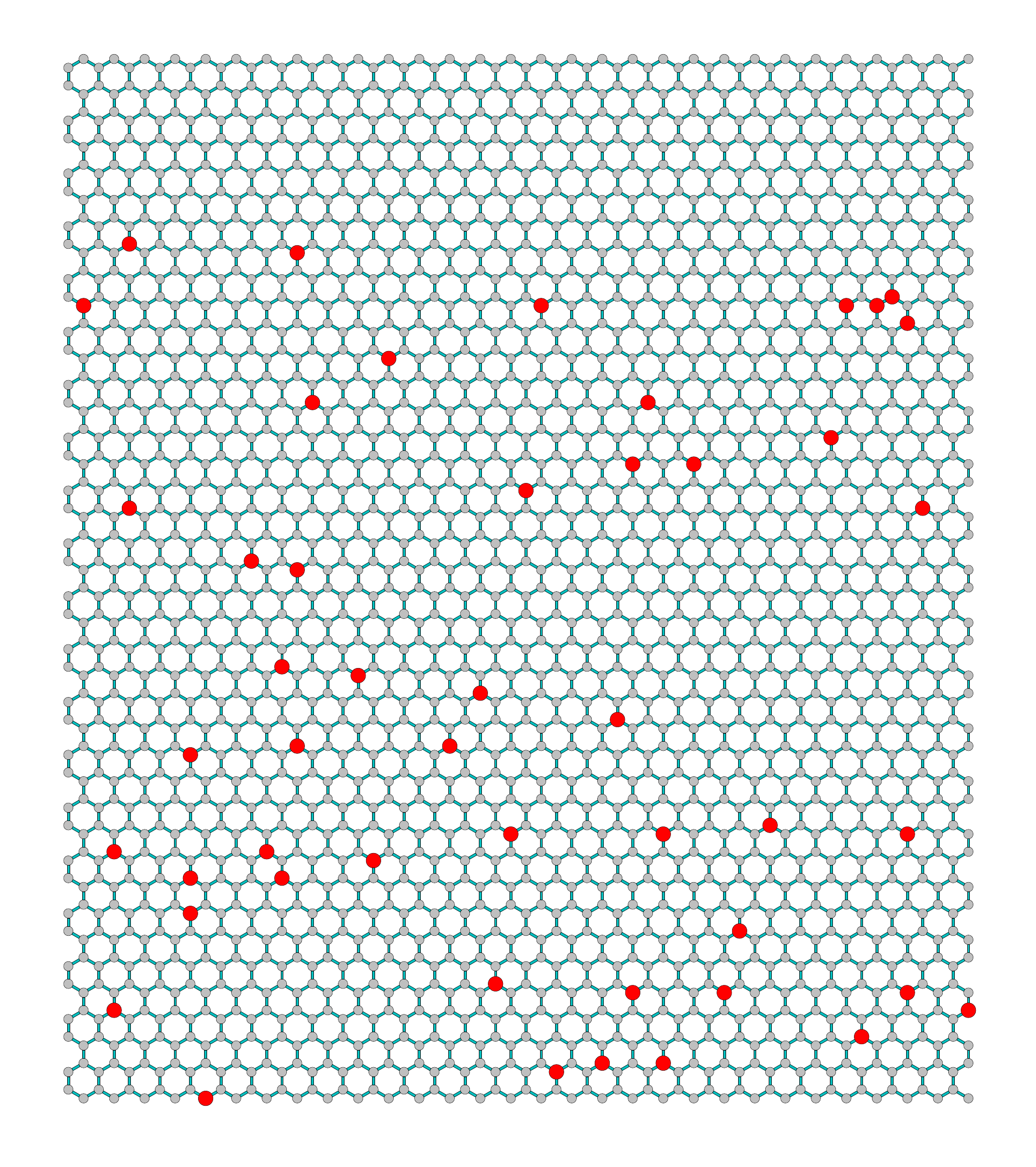}
}
\mbox{
\includegraphics[width=4cm]{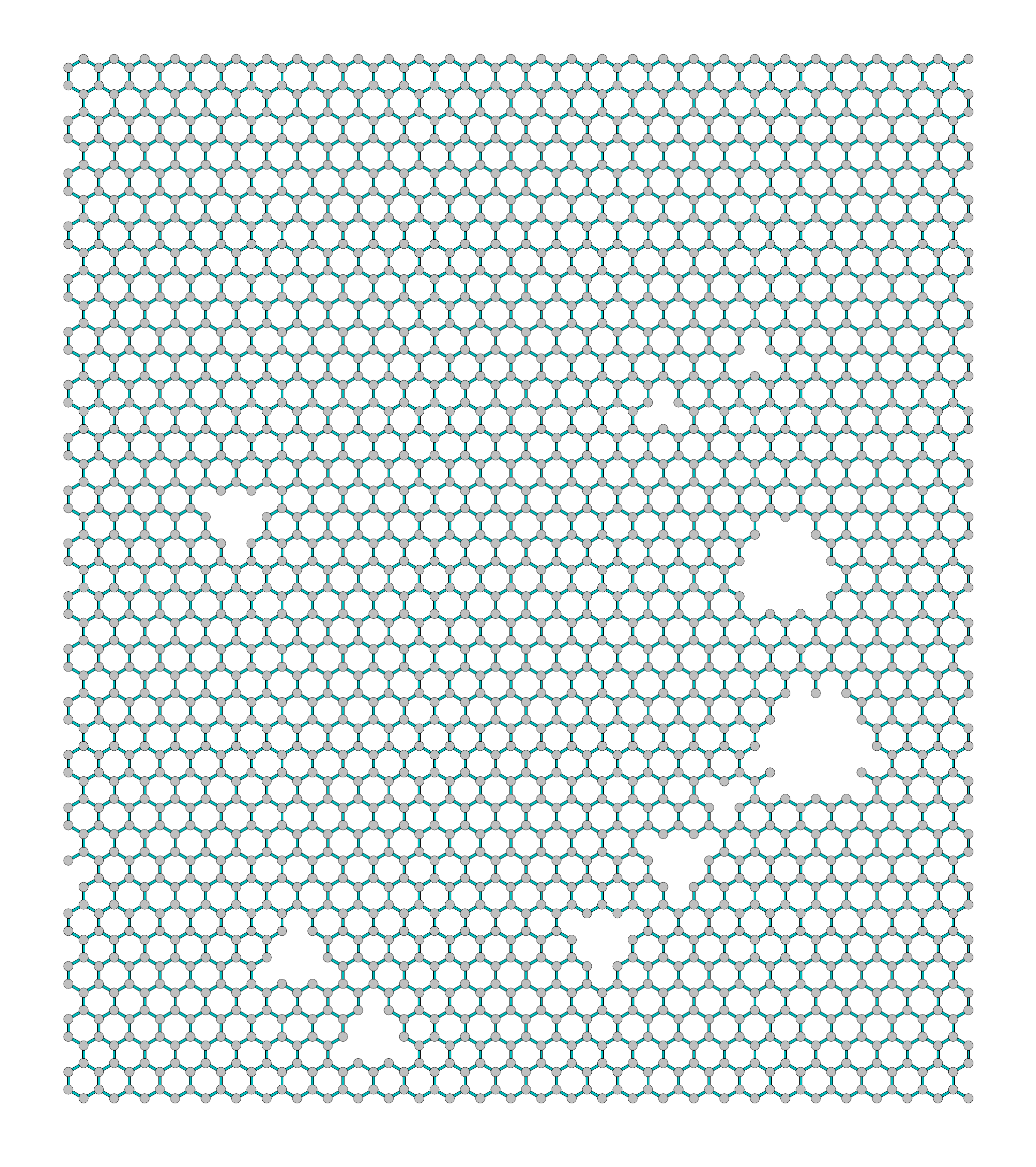}
\includegraphics[width=4cm]{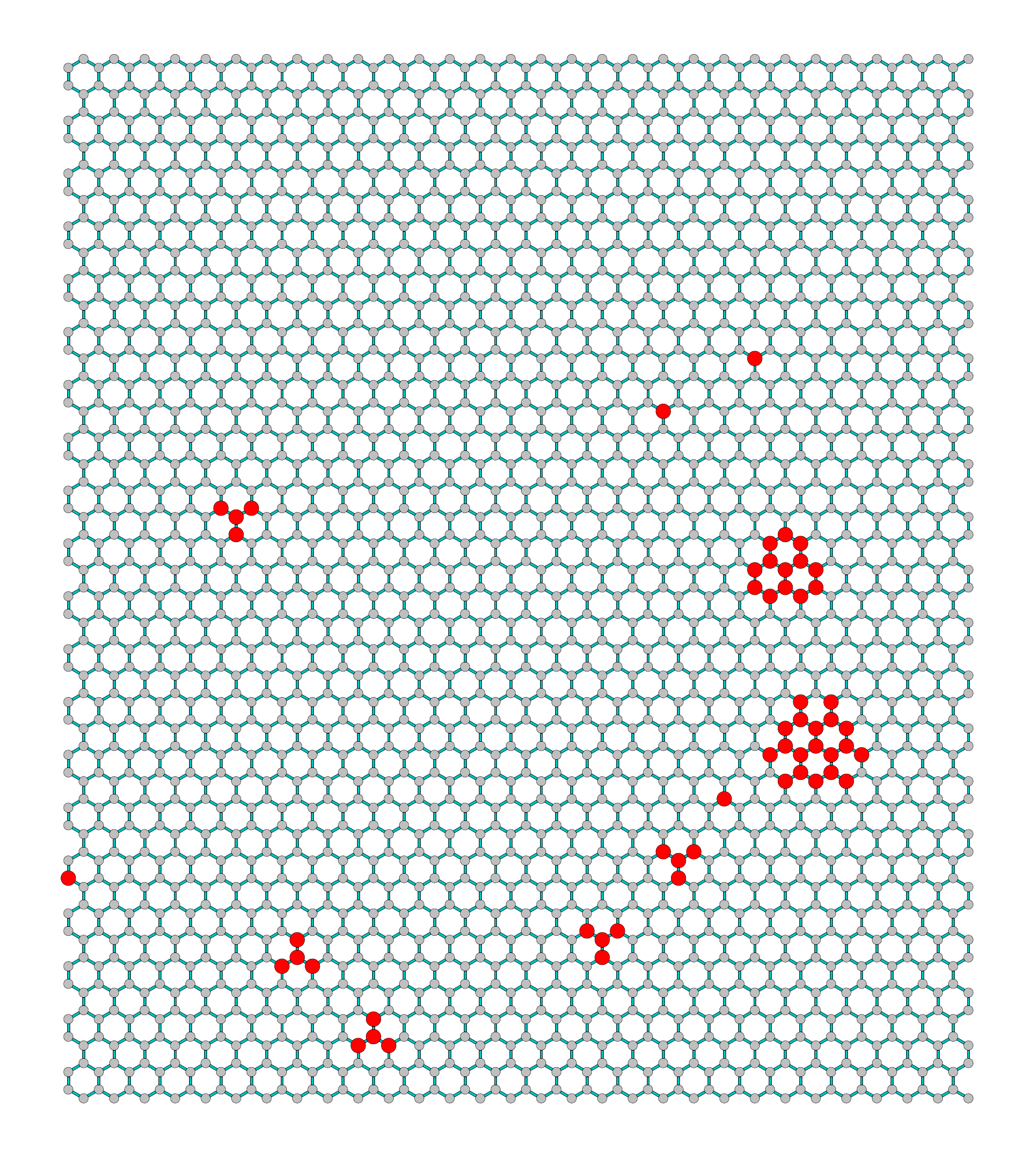}
}
\mbox{
\includegraphics[width=4cm]{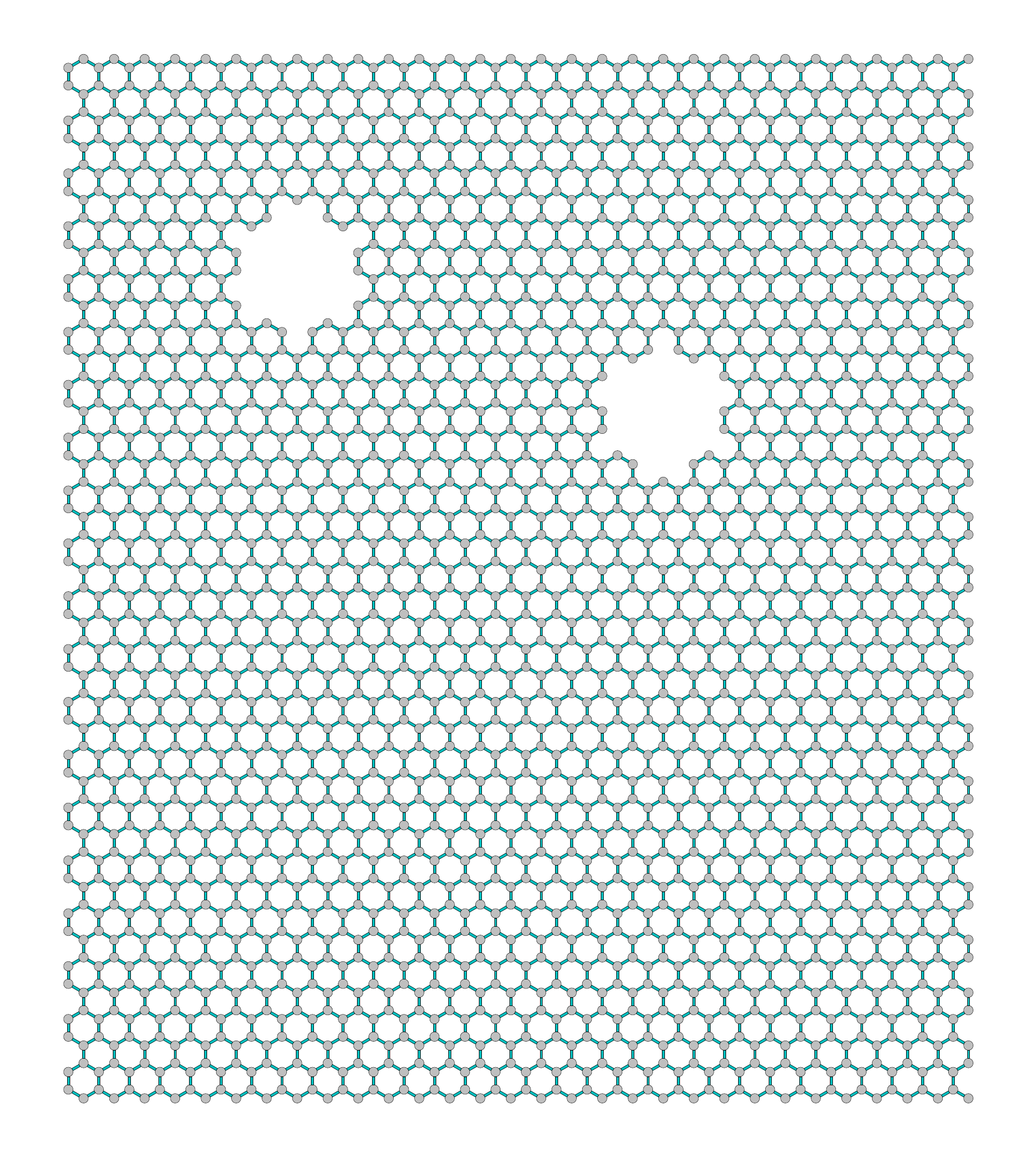}
\includegraphics[width=4cm]{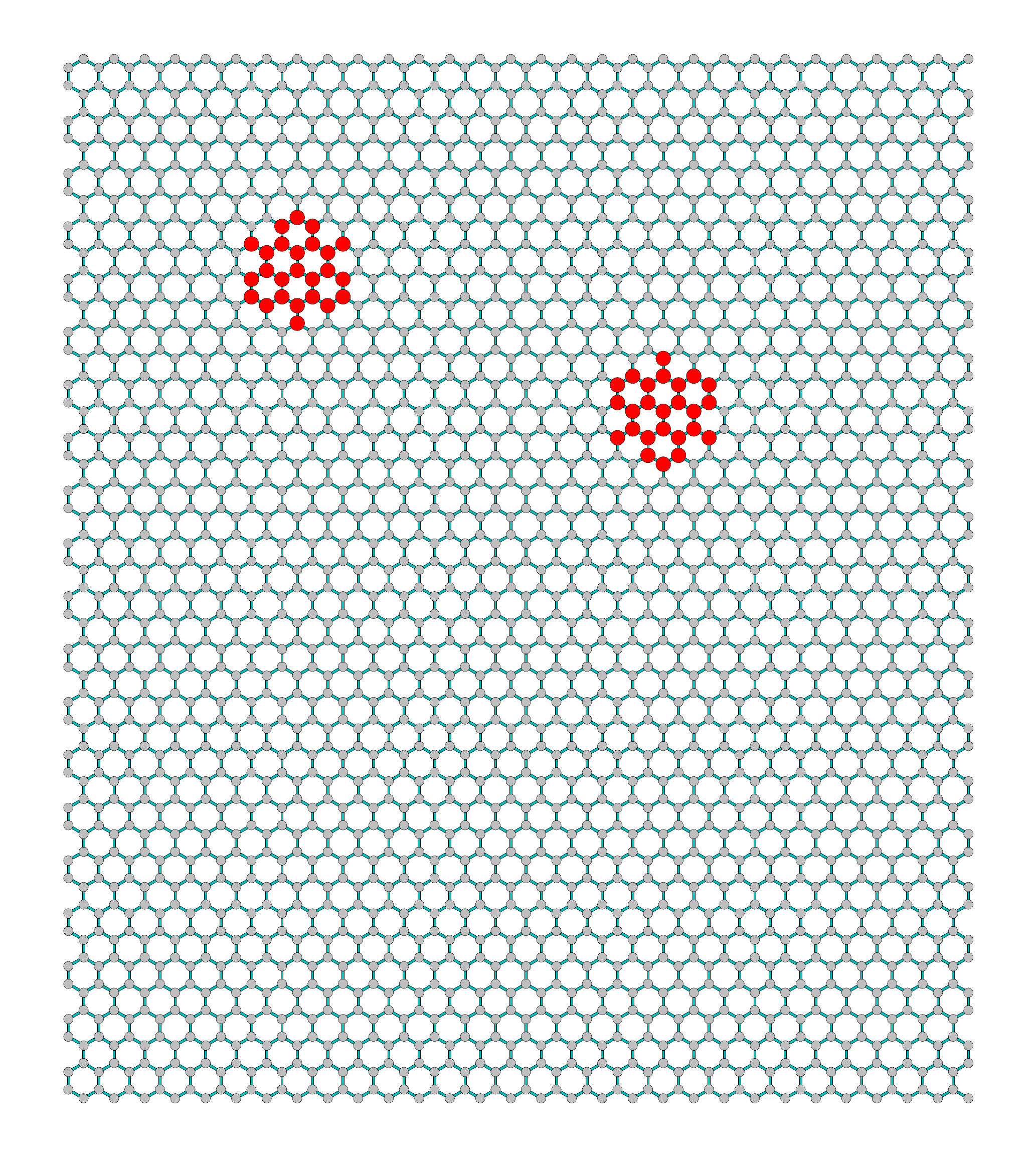}
}
\end{center}
\caption{(Color online) Sketch of a gaphene sheet with vacancies (left
pannels) or hydrogen adatoms (right pannels). The vacancies are presented
as missing carbon atoms, whereas the hydrogen adatoms are
highlighted in red. From top to bottom: Resonant impurites are
distributed as the formation I ($R=0)$, II ($0\leq R\leq 3a$) and III ($R=3a$%
) as described in the text. For illustrative purposes, the size of the sample
shown in this sketch is $60\times 40$, and the concentration of
impurities is approximately equal to $2\%$.}
\label{figcluster}
\end{figure}

Correlated resonant impurities are introduced by the formation of groups of
vacancies or adsorbed hydrogen atoms (see Fig. \ref{figcluster}). The center
of the formed vacancy or hydrogen cluster ($\mathbf{r}_{c}$) is randomly
distributed over the honeycomb lattice sites, with equal probability on both
sublattices A and B. Each site ($i$) whose distance to one of the centers ($R\equiv
\left\vert \mathbf{r-r}_{c}\right\vert $) is smaller than a certain value ($%
R_{c}$), is assumed to be part of the cluster, i.e., being a vacancy or
adsorbing a hydrogen atom. We further introduce another freedom of the
resonant clusters namely that their radius can change within the sample,
allowing for a graphene layer with cluster of impurities of different size.
This means that the value of $R_{c}$ for each resonant cluster can either be
different and randomly distributed to a maximum value, or can be kept fixed
for all the clusters in the sample. We want to emphasize that as
the center of the cluster is located on a particular sublattice A or B, the formation
of the cluster does not preserve the sublattice symmetry and therefore can lead
to the appearance of midgap states.
\begin{figure*}[t]
\begin{center}
\mbox{
\includegraphics[width=7cm]{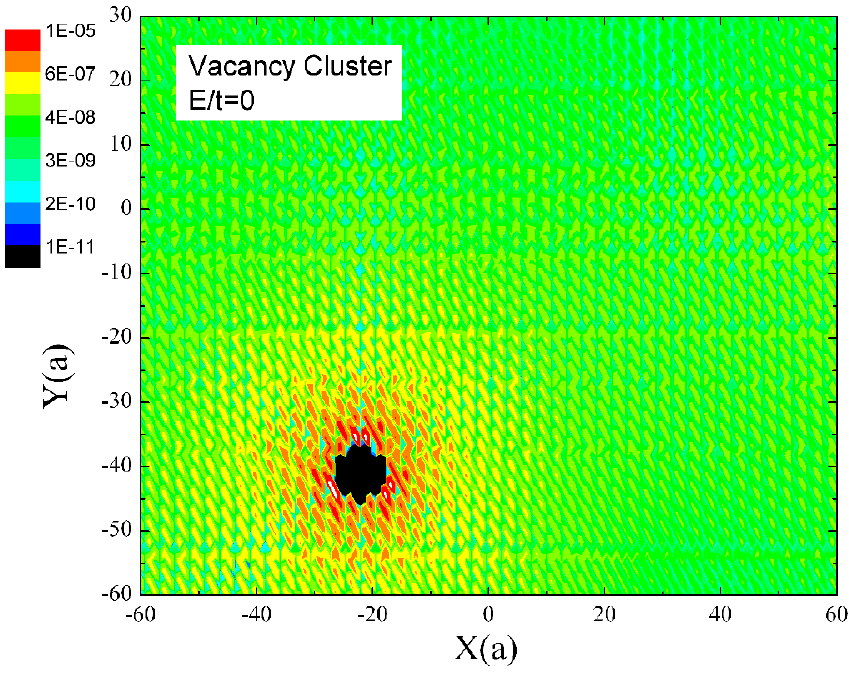}
\includegraphics[width=7cm]{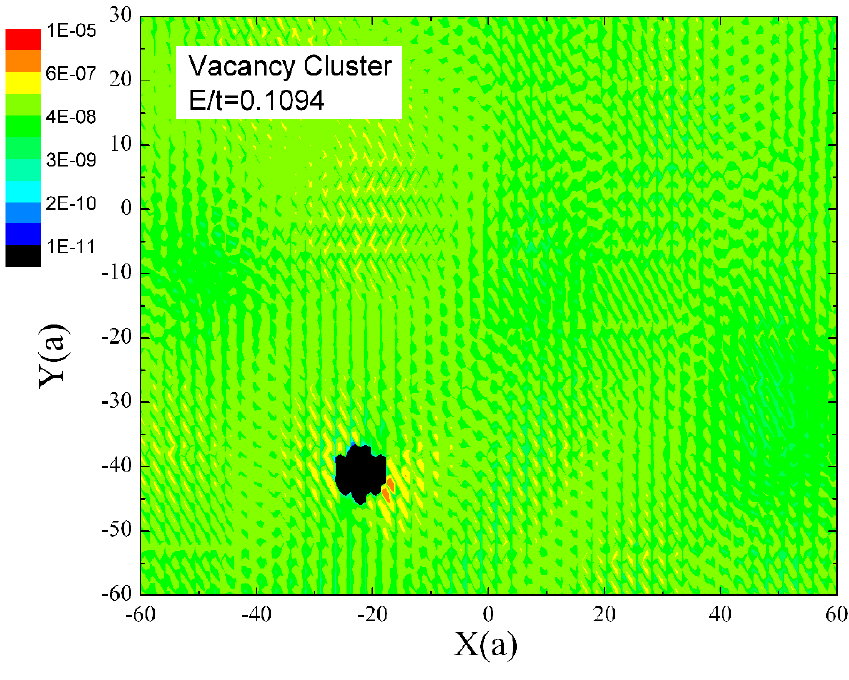}
}
\mbox{
\includegraphics[width=7cm]{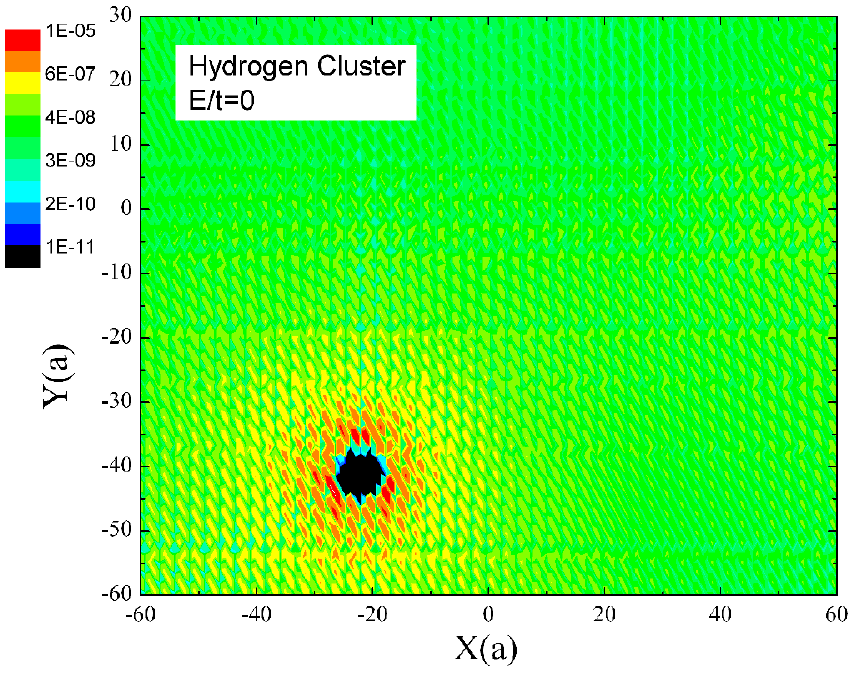}
\includegraphics[width=7cm]{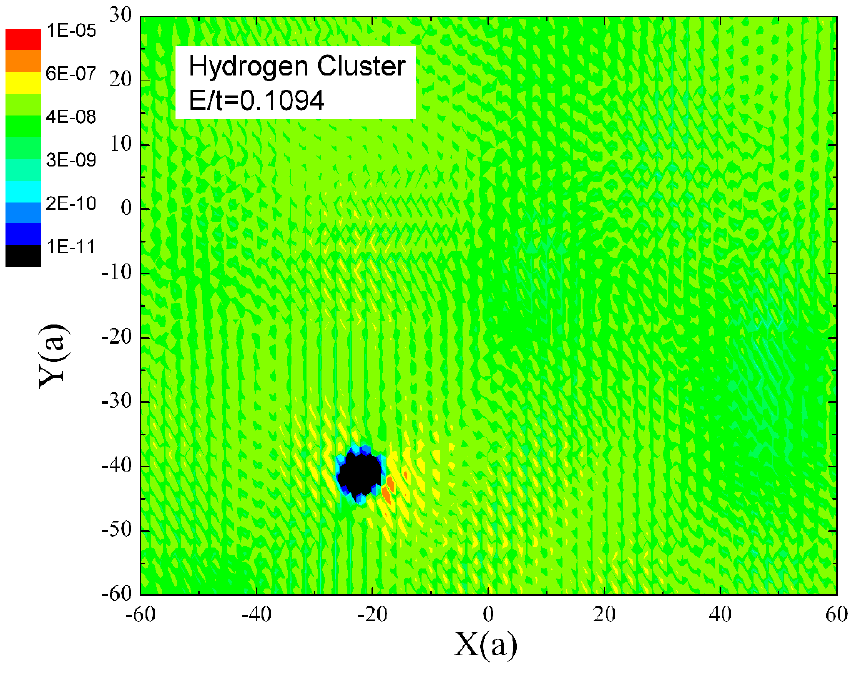}
}
\end{center}
\caption{(Color online) Contour plot of the amplitudes of quasieigenstates
at energy $E=0$ or $E=0.1094t$. The radius of the resonant clusters is fixed
at $R_{c}=5a$.}
\label{quasieigenstates}
\end{figure*}

Firts, in Fig. \ref{dos_ac_cluster}(e) and (g), we compare the
density of states with the same total number of resonant
impurities (vacancies or hydrogen adatoms) but with different
kinds of formations. We consider three different situations, i.e.,
randomly distributed uncorrelated single impurities (formation I),
randomly distributed correlated clusters with varied radius of
clusters (formation II) or with fixed radius of clusters
(formation III). The different structures are sketched in Fig. \ref%
{figcluster}. Notice that the formation I is a limiting case of the
formation III with all the radius of clusters being zero. As we can see from
the results of the simulations, the number of midgap states is larger in the
cases of uncorrelated single resonant impurities, and smaller for the case
of fixed radius of resonant clusters. This is expected since the midgap
states are states which are quasilocalized around the vacancies or carbon
atoms which adsorb hydrogen atoms.\cite%
{Peres2006,Pereira2006,Pereira2008,YRK10} Therefore, for the
same concentration of impurities, the number of midgap states will grow with
the \textit{isolation} of the impurities in small clusters. Something
similar happens for the case of hydrogen clusters. This can understood by
looking at Fig. \ref{quasieigenstates}, where we present contour plots of the amplitudes
of quasieigenstates at the Dirac point or outside the midgap region. The
quasieigenstate $\left\vert \Psi \left( \varepsilon \right) \right\rangle $
is a superposition of the degenerated eigenstates with the same eigenenergy $%
\varepsilon $, obtained by the Fourier transformation of the wave functions
at different times\cite{YRK10}%
\begin{equation}
\left\vert \Psi \left( \varepsilon \right) \right\rangle =\frac{1}{2\pi }%
\int_{-\infty }^{\infty }dte^{i\varepsilon t}\left\vert \varphi \left(
t\right) \right\rangle ,
\end{equation}%
where $\left\vert \varphi \left( t\right) \right\rangle =e^{-iHt}\left\vert
\varphi \right\rangle $ is the time evolution of the initial state $%
\left\vert \varphi \right\rangle $ defined in Eq.~(\ref{Eq:phi0}). Although
the quasieigenstate is not exactly the energy eigenstate unless the
corresponding eigenstate is not degenerated at energy $\varepsilon $, we can
still use the distribution of the amplitude in the real space to verify the
quasilocalization of the zero modes in the presence of random impurities,
\cite{YRK10} or obtain the DC conductivity at certain energies or carrier
densities.\cite{YRK10,WK10,YRK10b} As we can see from Fig. \ref%
{quasieigenstates}, the contour plots of the quasieigenstates of graphene
with vacancy and hydrogen clusters are quite similar, i.e., the amplitudes
on the carbon atoms which adsorb an hydrogen atom are almost zero, just like
if they are vacancies. Furthermore, at the Dirac point (left
panels of Fig. \ref{quasieigenstates}, corresponding to $E=0$) the
quasieigenstates are semi-localized around the edge of the clusters (see the
red color in the regions around the cluster). On the other hand, for
energies above the impurity band, the states are not localized around the
resonant cluster, and the amplitudes of the quasieigenstates are more or
less uniformly distributed over the sample except within the clusters, where
the amplitudes are zero. Therefore, as we have discussed above, for a given
concentration of impurities, the number of carbon atoms which are located
around an impurity will be larger in the formation I than in the formation
III. Then, the number of zero modes is also larger in I than in III, leading
to spectra for the DOS and optical conductivity similar to the ones of clean
graphene for samples in which disorder is concentrated in a small number of
big clusters (formation III) than spread into a large number of small
clusters (formation I), as it can be seen in Figs. \ref{dos_ac_cluster}%
(e)-(h). Finally, notice that the possibility for new excitations between
the impurity and the carrier bands, leads to a modulation of the optical
conductivity (as compared to the clean membrane) whose peak structure
depends on the renormalized DOS and band dispersion of each case.

\section{Optical conductivity of doped graphene}

\label{Sec:Doped}

\begin{figure}[t]
\begin{center}
\mbox{
\includegraphics[width=7cm]{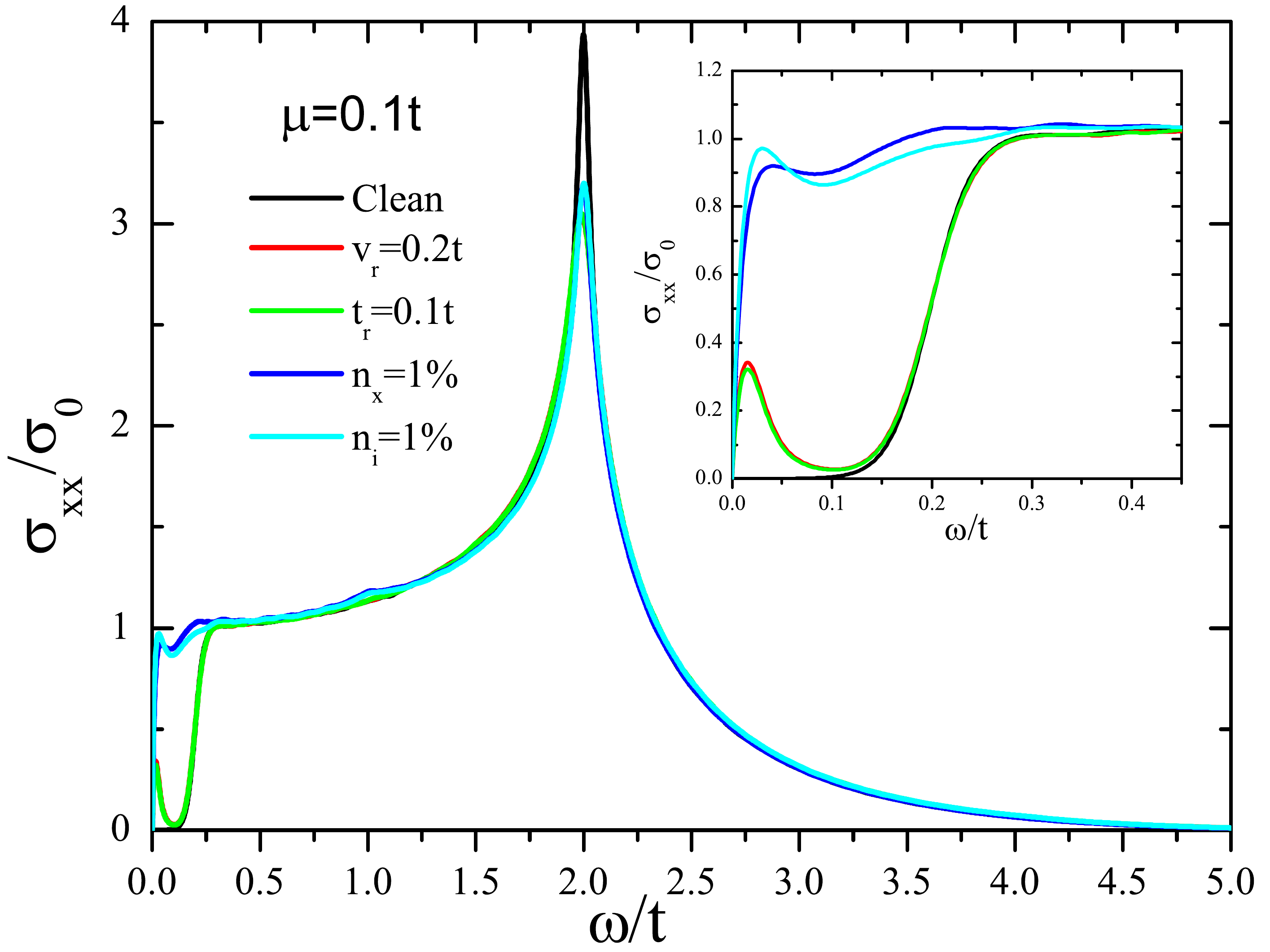}
} \mbox{
\includegraphics[width=7cm]{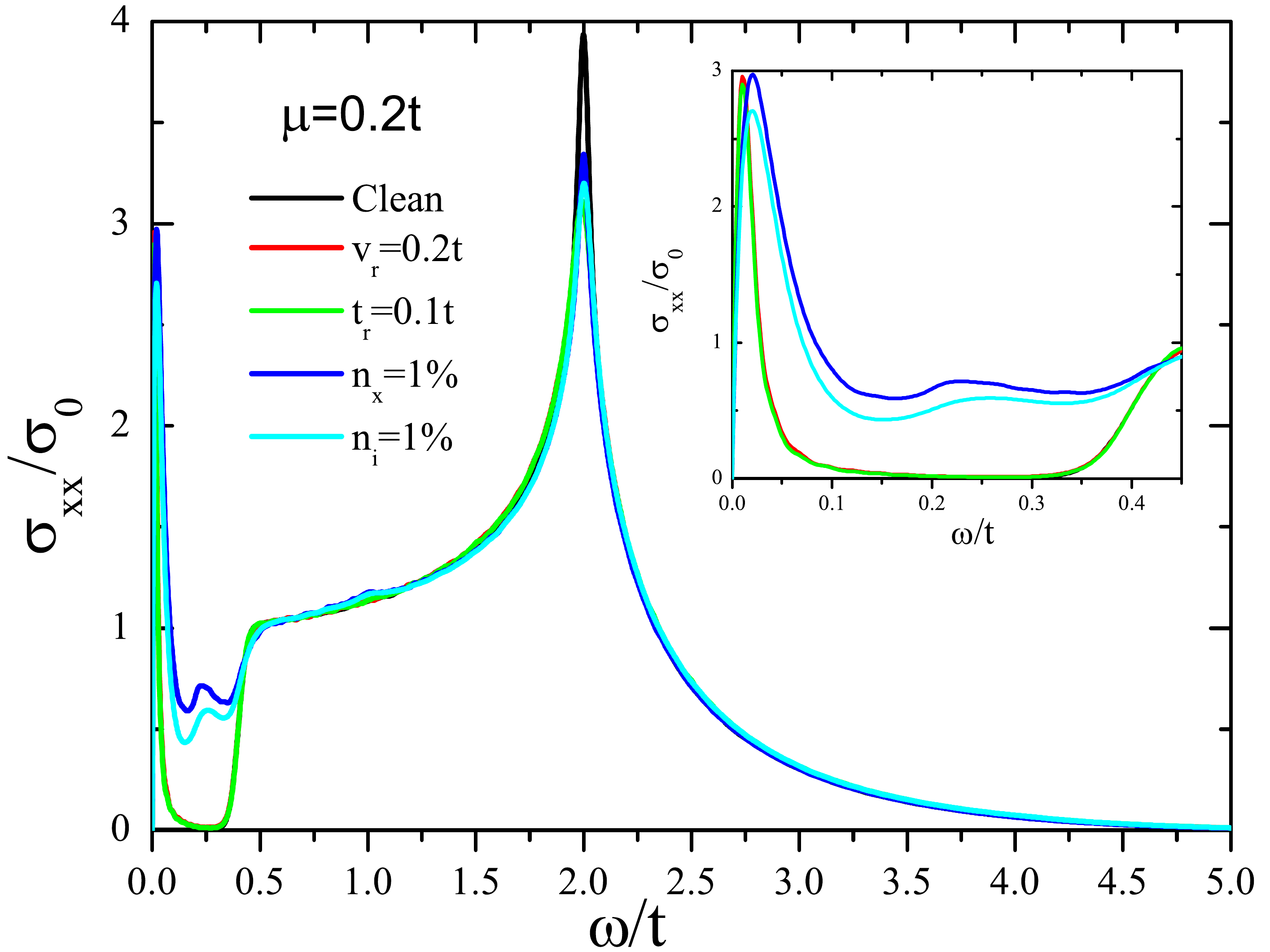}
}
\end{center}
\caption{(Color online) Simulation results of the optical conductivity of
doped graphene with diffenrent kinds of non-correlated disorders. The
chemical potential is $\protect\mu =0.1t$ in (a) and $0.2t$ in (b).}
\label{ac_slg_miu}
\end{figure}

So far we have discussed the effects of disorder on the optical response of
undoped graphene. In this section, we study the optical conductivity of
graphene for finite values of the chemical potential, taking into account
the effect of disorder. At zero temperature, a clean sheet of gated (doped)
graphene has a zero optical conductivity in the region $\omega <2\mu $, and
an universal conductivity of $\sigma(\omega)=\sigma _{0}$, due to optically
active inter-band excitations through the Dirac point, for energies above
the threshold $\omega>2\mu $.\cite%
{Ando2002,Stauber2008,Falkovsky2007,Kuzmenko2008,YRK10} In the presence of the
disorder, the broadening of the bands as well as the appearance of possible
midgap states leads to a more complicated selection rule for the optical
transitions, making possible to have excitations in the \textit{forbidden
region} $0<\omega<2\mu$, as observed experimentally.\cite{LB08}
In this section, we are interested in studying the effect
on the optical spectrum of doped graphene of the different
kinds of disorder, considered in the previous section.

\begin{figure*}[t]
\begin{center}
\mbox{
\includegraphics[width=7cm]{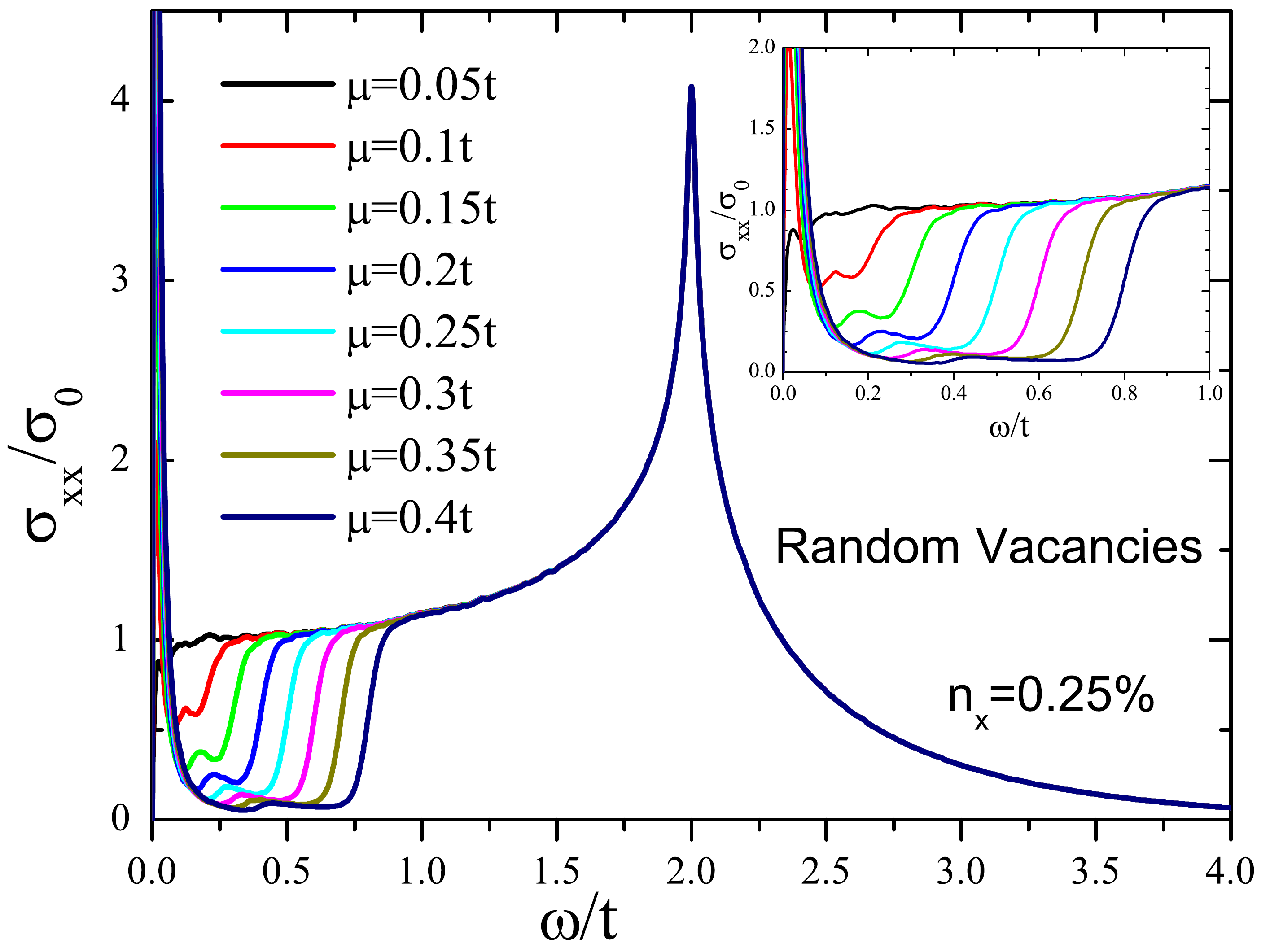}
\includegraphics[width=7cm]{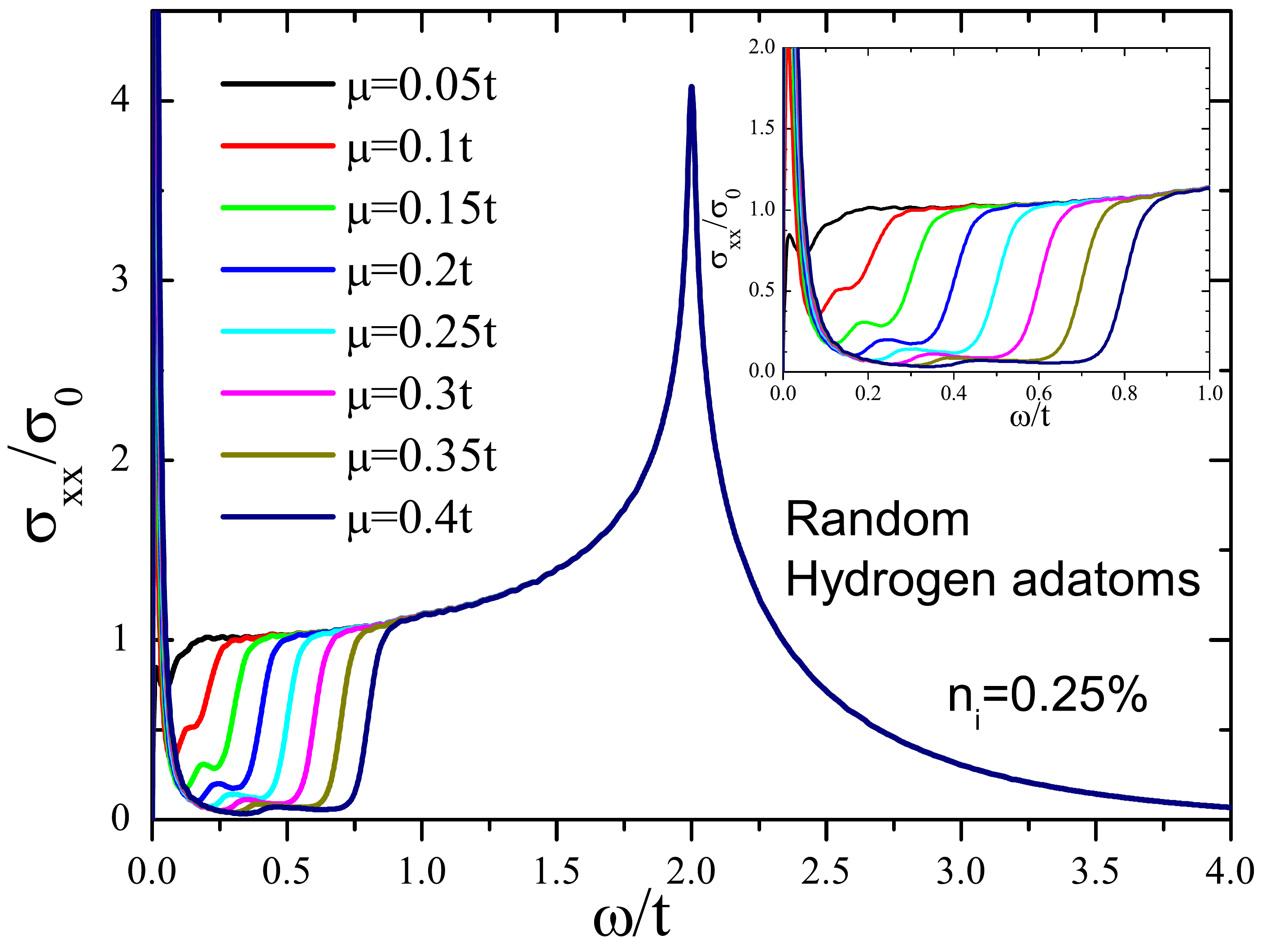}
}
\mbox{
\includegraphics[width=7cm]{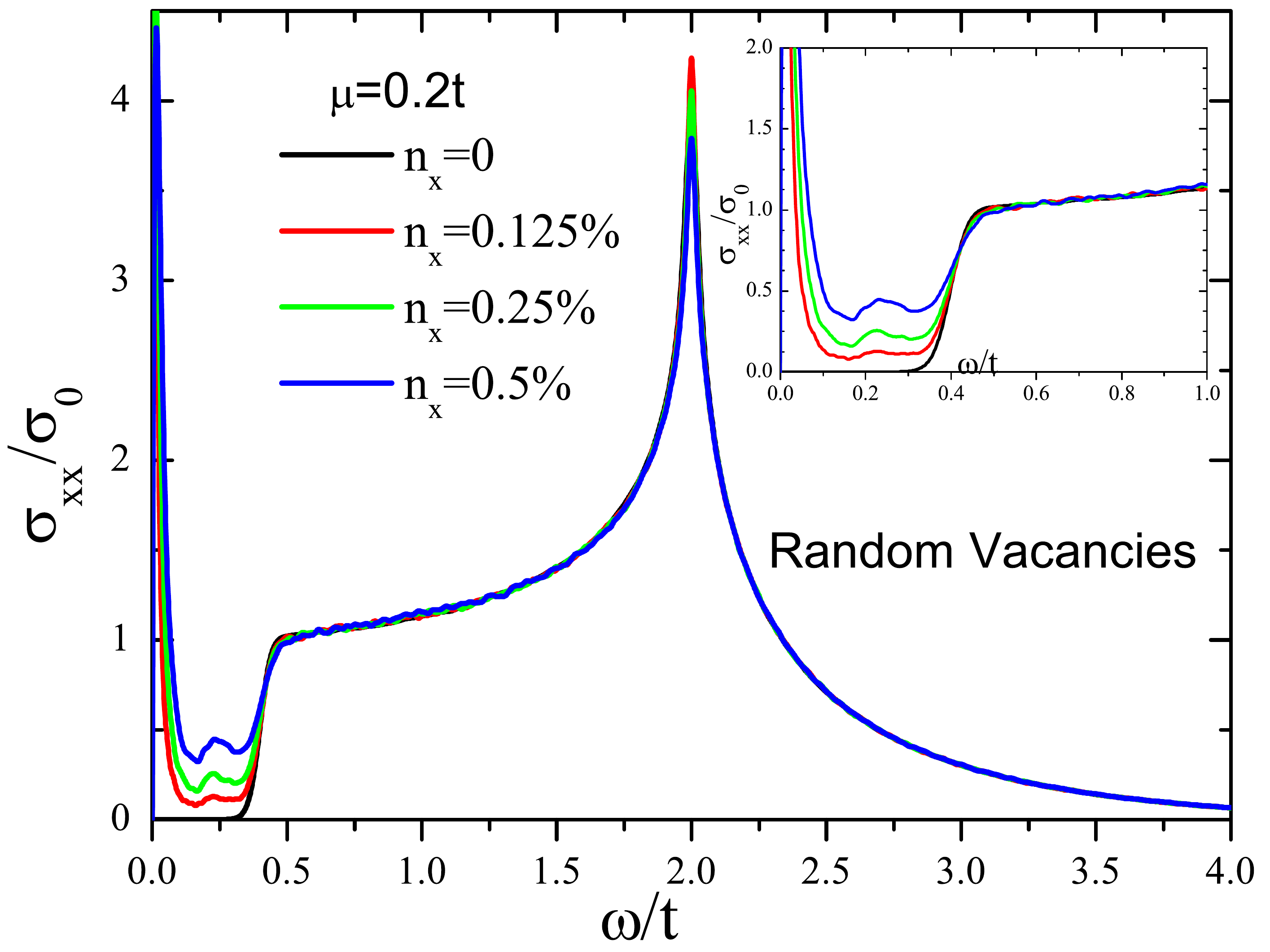}
\includegraphics[width=7cm]{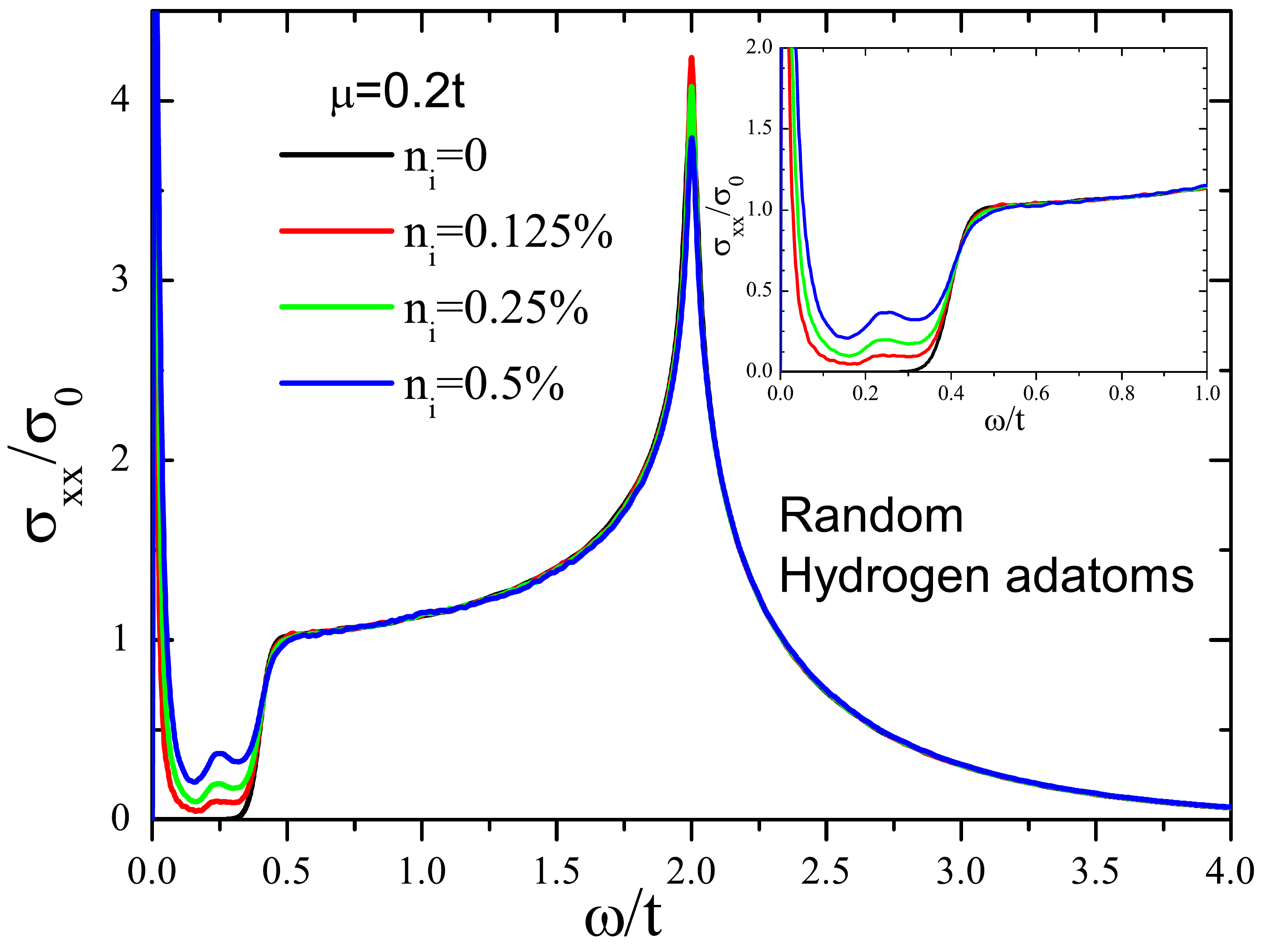}
}
\end{center}
\caption{(Color online) Optical conductivity of doped graphene with resonant
scatterers. Upper panels: Fixed concentration of impurities, $\protect%
\sigma(\protect\omega)$ for different values of $\protect\mu$. Lower panels:
Fixed chemical potential $\protect\mu$, $\protect\sigma(\protect\omega)
$ for different concentration of impurities.
}
\label{Fig:Doped-resonant}
\end{figure*}

In Fig. \ref{ac_slg_miu} we compare the numerical results for the
optical conductivity of doped graphene, considering four different
types of non-correlated disorder (random potentials, random
hoppings, vacancies and hydrogen adatoms) as well as clean
graphene. First, one notice that the effect of doping is not
relevant in the high energy part of the spectrum ($\omega\gg
\mu$), and $\sigma(\omega)$ follows the same behavior discussed in
Secs. \ref{Sec:NCD} and \ref{Sec:CD}, with a peak corresponding to
particle-hole inter-band transitions between states of the Van
Hove singularities at $\omega\approx 2t$. However, the spectrum
changes dramatically in the infra-red region, as shown in the insets
of Fig. \ref {ac_slg_miu}. Therefore, from now on we will focus
our interest on the effect of disorder on this low energy part of
the spectrum. First, one notices that for all kinds of disorder,
there is a peak in $\sigma(\omega)$ close to $\omega=0$, whereas
at slightly higher energies, $\sigma(\omega)$ drops to almost zero
for the case of non-resonant scatterers (red and green curves),
while there is still a non-zero background contribution when
resonant scatterers are considered (light and dark blue curves).
This can be understood as follows: for all the cases, disorder
leads to a broadening of the bands, which allows for intra-band
transitions between surrounding states of the Fermi level.
However, we have seen that resonant impurities create an impurity
band at the Dirac point, with the corresponding peak in the DOS at
$E=0 $, whereas non-resonant impurities are not so effective
in creating midgap states. Therefore, the background contribution
that we find in Fig. \ref {ac_slg_miu}(a)-(b) between
$0<\omega<2\mu$ for samples with resonant scatterers are due to
transitions between the newly formed impurity band and the
conduction band. Taking into account that resonant impurities are
believed to be the main source of scattering in
graphene,\cite{WK10,MM10,NG10} our results suggest that this kind
of impurities could be behind the background contribution to the
optical conductivity observed
experimentally.\cite{LB08,Mak2008,ChenCF2011,Horng2011} Finally,
notice that the peak observed in $\sigma(\omega)$ for the case of
resonant impurities at the energy $ \omega\approx \mu$ is
associated to transitions between the above discussed
impurity band and states at the Fermi level.

\begin{figure*}[t]
\begin{center}
\mbox{
\includegraphics[width=7cm]{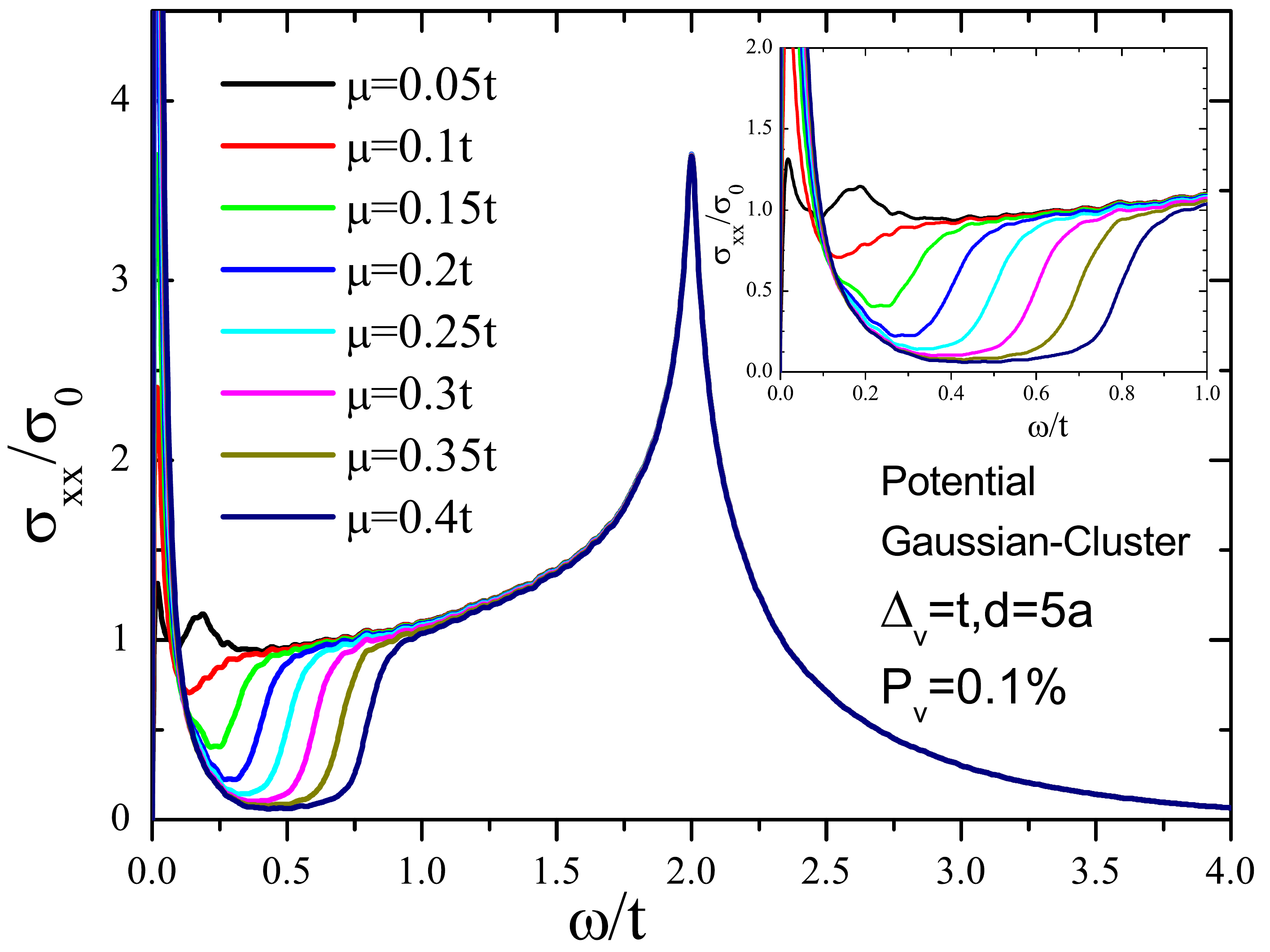}
\includegraphics[width=7cm]{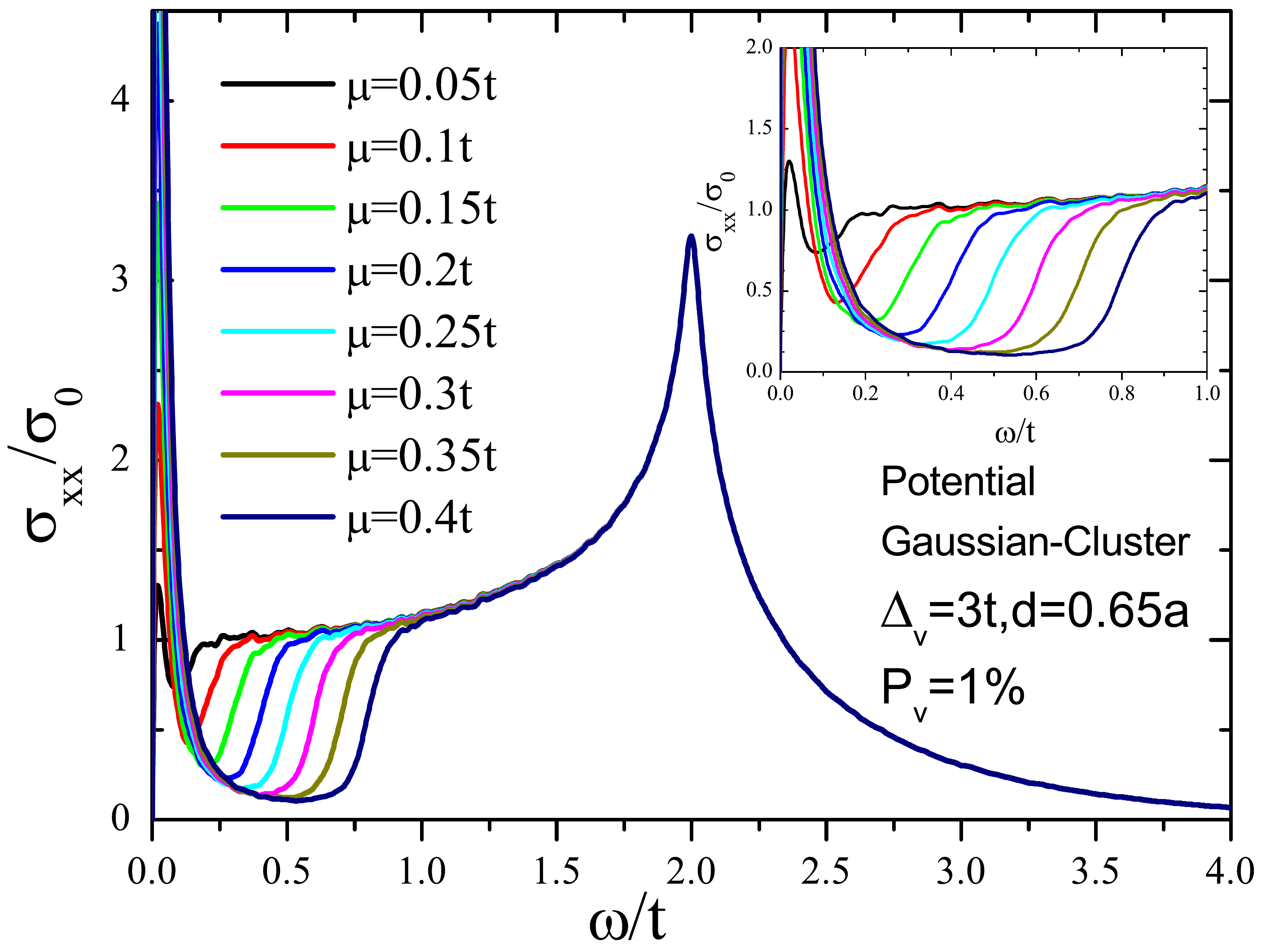}
}
\mbox{
\includegraphics[width=7cm]{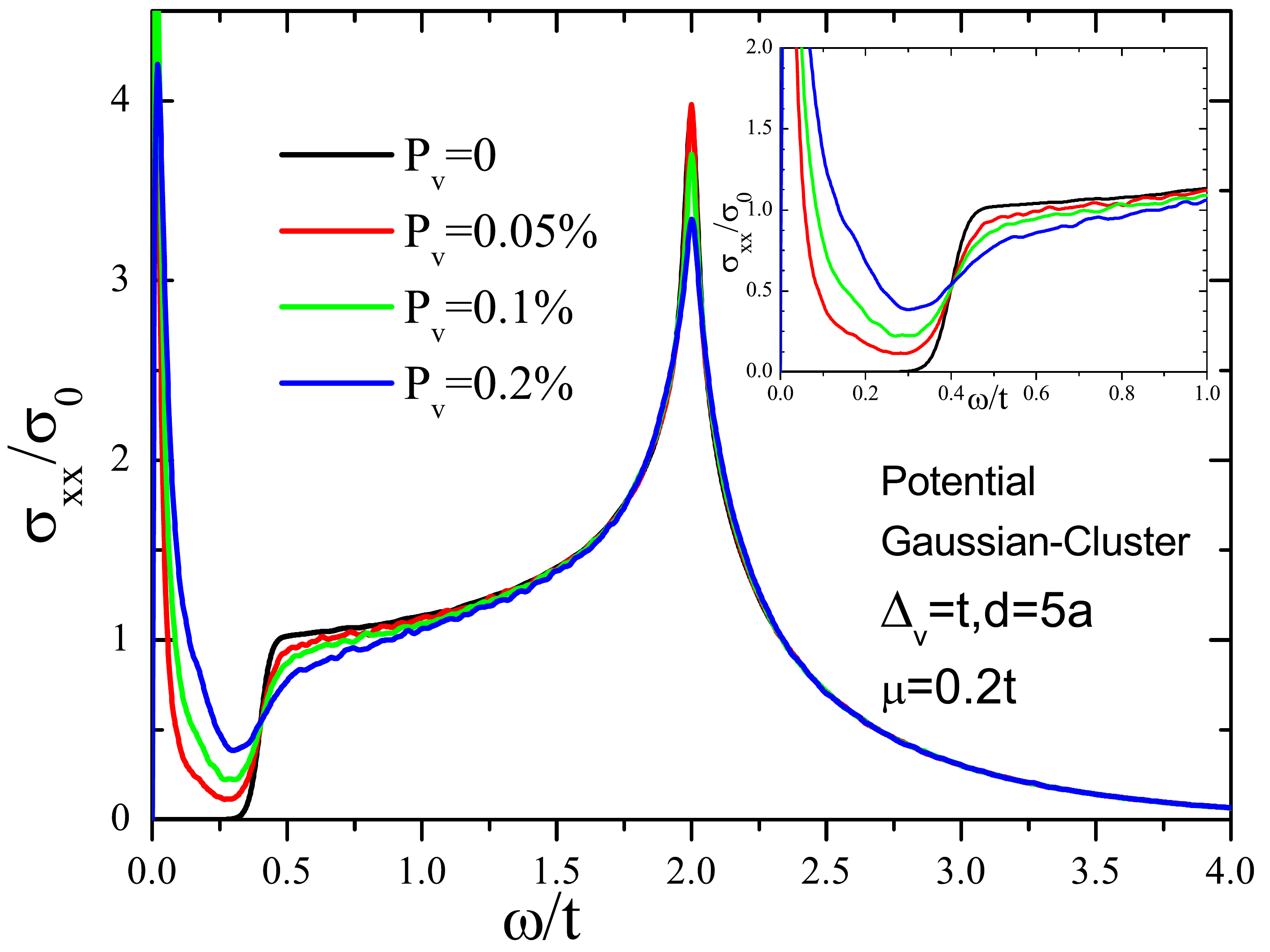}
\includegraphics[width=7cm]{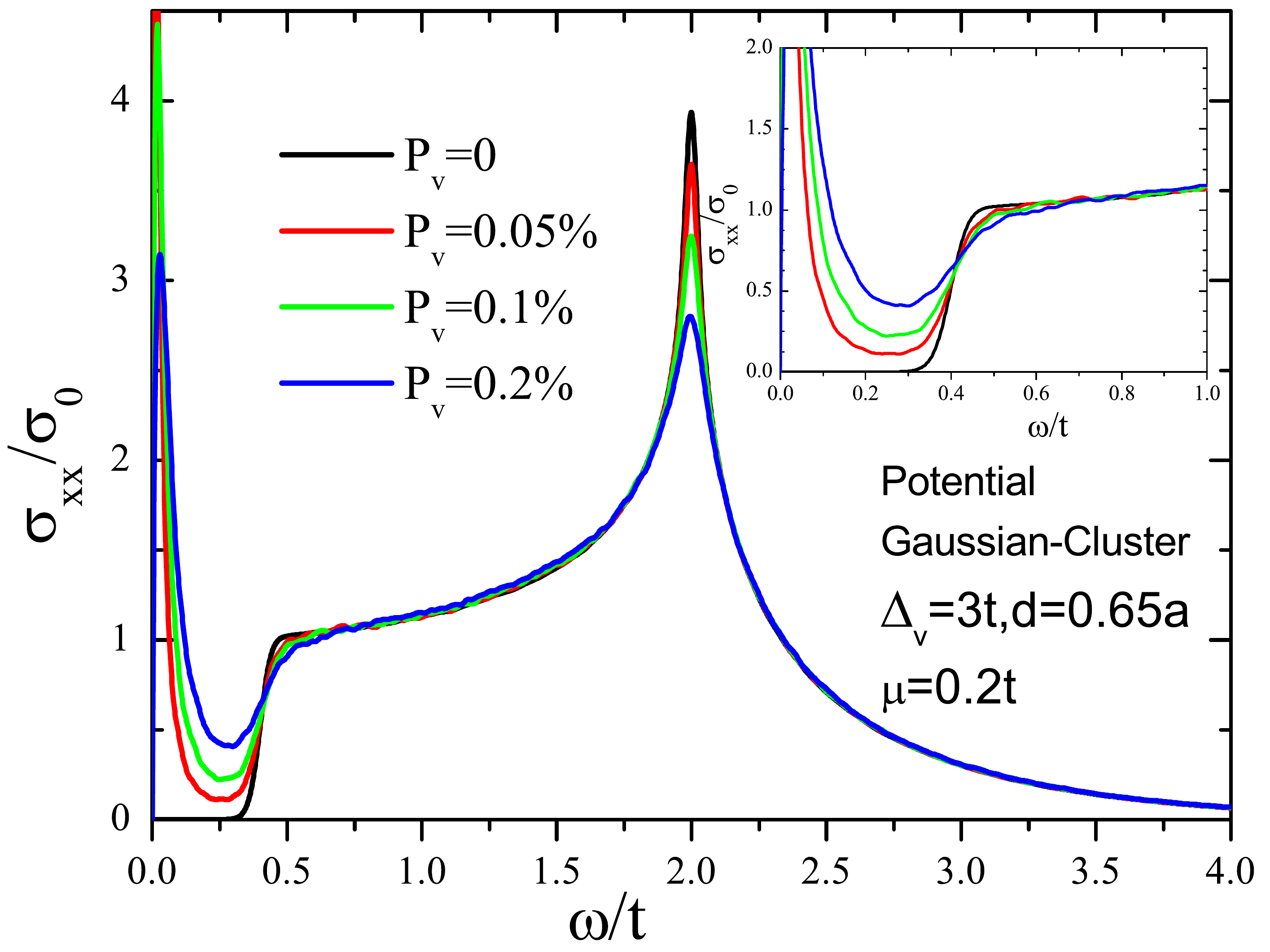}
}
\end{center}
\caption{(Color online) Same as Fig. \protect\ref{Fig:Doped-resonant} but
for doped graphene with on-site potential disorder distributed in gaussian
clusters.
}
\label{Fig:Doped-Gaussian}
\end{figure*}

To gain more insight about the effect of disorder in the optical
conductivity of doped graphene, in Fig. \ref{Fig:Doped-resonant} we show $%
\sigma (\omega )$ for different values of $\mu $ at fixed
concentration of impurities (upper panels), and $\sigma (\omega )$
for different concentrations of impurities and fixed $\mu $. In
the first case, the main feature is that the conductivity
increases as the doping decreases, in a qualitative agreement with
the experimental results.\cite{LB08} When the chemical potential
is fixed and the concentration of impurities
changes (bottom panels), one observe that the conductivity in the region $%
0<\omega <2\mu $ grows with $n_{x(i)}$ from $\sigma (\omega )=0$ for a clean
sample to $\sigma (\omega )\approx 0.4\sigma _{0}$ for the larger
concentration of impurities consider ($n_{x(i)}=0.5\%$). If we compare to
recent experiments, we notice that a $0.25\%$ of resonant impurities would
lead to a background contribution similar to the one reported by Li \textit{%
et al.} for graphene on SiO$_{2}$,\cite{LB08} whereas only a $\sim
0.1\%$ of resonant impurities would be necessary to quantitatively
reproduce the results of Chen \textit{et al.} for graphene doped
with a high-capacitance ion-gel gate dielectric.\cite{ChenCF2011}
Finally, we can see that similar results are
obtained when a sample with correlated on-site potential disorder
distributed in the form of Gaussian clusters, as shown in
Fig. \ref{Fig:Doped-Gaussian}. Therefore, we conclude that there
are several kinds of disorder (resonant scatterers and correlated
impurities) that can induce a finite conductivity in the infra-red
region of the spectrum, as observed experimentally. It
is the whole set of data on DC and AC transport from which
one may infer the dominant type of defects in real graphene.

\section{Conclusion and Discussion}

\label{Sec:Conclusions}

We have presented a detailed theoretical study of
the optical conductivity of graphene with different kinds of
disorder, as resonant impurities, random distribution of on-site
potential or random renormalization of the nearest-neighbor
hopping parameter (which can account for the effect of
substitutional defects). Furthermore, we have consider the
possibility for the impurities to be correlated or non-correlated.

For all types of disorder considered, the high energy peak at
$\omega\approx 2t$, due to inter-band excitations between states
of the Van Hove singularities of the valence and the conduction
bands, are always sensitive to disorder, getting smeared out
proportionally to the strength of disorder. On the other hand, the
low energy part of the optical spectrum is strongly dependent on
the type of disorder, as well as its strength and concentration.
In general, for undoped graphene and in the presence of small
disorder of the on-site potentials or in the nearest-neighbor
hopping between the carbon atoms, the characteristics of the
single-particle Dirac
cone approximation are clearly present in the spectrum, and $\sigma(\omega)%
\approx\sigma_0$ at energies for which the continuum approximation
applies. This is also true when we consider Gaussian hopping
parameters. On the other hand, if there are long-range
Gaussian potentials, the local shifts of the Dirac points leads to
electron-hole puddles and to the emergence of states in the
vicinal region of the Dirac points. As a consequence, we observe
an enhancement of the optical conductivity in the infra-red part
of the spectrum. Interestingly, in the presence of resonant
impurities (vacancies or hydrogen adatoms) there appear midgap
states which are quasilocalized around the impurities, the number
of which is proportional to the number of carbon atoms which are
located around the impurities. Completely random distributed
(non-correlated) resonant impurities lead to the strongest
enhancement of zero modes (seen as a prominent peak in the DOS at
zero energy) and also the largest effect on the optical spectrum.
In fact, for a large enough amount of resonant impurities, we
obtain a new peak in the optical conductivity at
an energy $\omega\approx t$, which is associated to optical
transitions between the midgap states and states of the Van Hove
singularities. When, for a given concentration of impurities, they
merge together forming clusters, instead of staying uncorrelated,
the influence of disorder on the electronic properties becomes
smaller, especially if these clusters form large islands.

Finally, we have considered the effect of doping on the spectrum.
Whereas
for clean graphene, only inter-band processes with an energy larger than $%
\omega=2\mu$ are optically active, the presence of disorder leads to a low
energy peak in $\sigma(\omega)$ (associated to transitions near the Fermi
level) plus a possible spectral weight in the region $0<\omega<2\mu$ for
disorders that can create an impurity band at zero energy. Most importantly,
we have found that a small amount of resonant impurities $\sim 0.1 - 0.2\%$%
, leads to a background contribution to $\sigma(\omega)$ between $%
0<\omega<2\mu$ in qualitative and quantitative agreement with recent
spectroscopy measurements.

\section{Acknowledgement}

The authors thank useful discussions with E. Cappelluti and F. Guinea.
This research is supported by the Stichting Fundamenteel Onderzoek der Materie
(FOM), the Netherlands National Computing Facilities foundation (NCF),
the EU-India FP-7 collaboration under MONAMI and the grant CONSOLIDER CSD2007-00010.

\bibliographystyle{apsrev4-1}
\bibliography{BibliogrGrafeno}

\end{document}